%% file: Disc_pot.tex
\documentclass[aps,prd,12pt,preprint,superscriptaddress,nofootinbib,floatfix]{revtex4}
\usepackage{latexsym}
\usepackage{amsfonts}
\usepackage[latin1]{inputenc}
\usepackage{subfig}
\usepackage{graphicx}
\usepackage{mathrsfs}
\usepackage{amssymb}
\usepackage{amsmath}
\usepackage{verbatim}
\def\D0{\slash\!\!\!\!\!\!\!\!\!\:D0}

\begin{document}

\begin{flushleft}
{SHEP-09-19\\ DFTT 58/2009}\\
\today
\end{flushleft}
\vspace*{-0.25cm}

\title{$Z'$ discovery potential at the LHC\\[0.15cm]
 in the minimal $B-L$ extension of the standard model}

\author{Lorenzo Basso}
\affiliation{
School of Physics \& Astronomy, University of Southampton,\\
Highfield, Southampton SO17 1BJ, UK}
\affiliation{
Particle Physics Department, Rutherford Appleton Laboratory, \\Chilton,
Didcot, Oxon OX11 0QX, UK}
\author{Alexander Belyaev}
\affiliation{
School of Physics \& Astronomy, University of Southampton,\\
Highfield, Southampton SO17 1BJ, UK}
\affiliation{
Particle Physics Department, Rutherford Appleton Laboratory, \\Chilton,
Didcot, Oxon OX11 0QX, UK}
\author{Stefano Moretti}
\affiliation{
School of Physics \& Astronomy, University of Southampton,\\
Highfield, Southampton SO17 1BJ, UK}
\affiliation{
Particle Physics Department, Rutherford Appleton Laboratory, \\Chilton,
Didcot, Oxon OX11 0QX, UK}
\affiliation{Dipartimento di Fisica Teorica, Universit\`a di Torino, \\
Via Pietro Giuria 1, 10125 Torino, Italy}
\author{Giovanni Marco Pruna}
\affiliation{
School of Physics \& Astronomy, University of Southampton,\\
Highfield, Southampton SO17 1BJ, UK}
\affiliation{
Particle Physics Department, Rutherford Appleton Laboratory, \\Chilton,
Didcot, Oxon OX11 0QX, UK}
\author{Claire H. Shepherd-Themistocleous}
\affiliation{
Particle Physics Department, Rutherford Appleton Laboratory, \\Chilton,
Didcot, Oxon OX11 0QX, UK}
\vspace*{-0.25cm}
\begin{abstract}
{\small \noindent
We present the Large Hadron Collider (LHC) discovery potential in the $Z'$ sector of a $U(1)_{B-L}$ enlarged Standard Model (that also includes
three heavy Majorana neutrinos and an additional Higgs boson)
for $\sqrt{s}=7$, $10$ and $14$ TeV centre-of-mass (CM) energies, considering both the $Z'_{B-L}\rightarrow e^+e^-$ and $Z'_{B-L}\rightarrow \mu^+\mu^-$ decay channels.
The comparison of the (irreducible) backgrounds with the expected backgrounds for the D$\O$ experiment at the Tevatron validates our simulation. We propose an alternative analysis 
that has the potential to improve the D$\O$ sensitivity.
Electrons provide a higher sensitivity to smaller couplings at small $Z'_{B-L}$ boson masses than do muons. The resolutions achievable may allow the $Z'_{B-L}$ boson width to be measured at smaller masses in the case of electrons in the final state.
The run of the LHC at $\sqrt{s}=7$ TeV, assuming at most $\int 
\mathcal{L} \sim 1$ fb$^{-1}$, will be able to give similar results to those that will be available soon at the Tevatron in the lower mass region, and to extend them for a heavier $M_{Z'}$. 
}
\end{abstract}
\maketitle

\newpage
\section{Introduction}
The evidence for non vanishing (although very small) neutrino masses is so far possibly the only hint for new physics beyond the Standard Model (SM) \cite{neutrino_mixing_angles/masses}. It is noteworthy that the accidental $U(1)_{B-L}$ global symmetry is not anomalous in the SM with massless neutrinos, but its origin is not well understood. It thus becomes appealing to extend the SM to explain simultaneously
the existence of both (i.e., neutrino masses and the $B-L$ global symmetry)
by gauging the $U(1)_{B-L}$
group thereby generating a $Z'$ state. This requires that the fermion and scalar spectra are enlarged to account for gauge anomaly cancellations. The results of direct searches constrain how this may be done \cite{Carena,Cacciapaglia:2006pk,Abazov:2010ti,Aaltonen:2011gp,D0}. Minimally, this requires the addition of a scalar singlet and three right-handed neutrinos, one per generation \cite{Buchmuller:1991ce,B-L}, which could dynamically trigger the see-saw mechanism explaining the smallness of the SM neutrino masses \cite{see-saw}. Within this model, the masses of the heavy neutrinos are such that their discovery falls within the LHC reach in a large portion of parameter space \cite{bbms,khalil-4l}.

Recently, a plethora of papers have been published studying the phenomenology of the $B-L$ model at colliders. They have dealt with the detectability of the $Z'$ at the LHC \cite{Coriano:2008wf,Emam:2007dy,Mine,bbms,Salvioni:2009mt} and at a future Linear Collider (LC) \cite{Freitas:2004hq,bbmp}, some analyses concentrating on the $Z'$ decaying via heavy neutrinos, in particular into three \cite{bbms} and four \cite{khalil-4l} leptons in the final state,  with distinctive displaced vertices due to long lived neutrinos, a clear signature of physics beyond the SM. Also, the testability at the LHC of the see-saw mechanism in this model has been evaluated in detail \cite{Perez:2009mu}.

In general, studies of this model focus on a specific non-disfavoured point in the parameter space and do not perform a systematic analysis of the entire space. The $Z'_{B-L}$ boson is also not always considered as a traditional benchmark for generic collider reach studies \cite{Gen_search,Carena,Coriano:2008wf,Ball:2007zza} or data analysis \cite{Abazov:2010ti,Aaltonen:2011gp,D0}. We have therefore performed a (parton level) discovery potential study for the LHC in the $Z'$ sector
of the $B-L$ model. In the light of the LHC plan of action for the next years \cite{LHC_cham}, we considered the CM energies of $7$ TeV and $14$ TeV, together with $10$ TeV as an eventual intermediate CM energy as originally planned. The integrated luminosities that we considered are up to $1$ fb$^{-1}$ for both $7$ TeV and $10$ TeV, and $100$ fb$^{-1}$ for $14$ TeV. We also included 
a comparison with the Tevatron reach for its expected $10$ fb$^{-1}$ of integrated luminosity. We chose to study the di-lepton channel (both electrons and muons), the cleanest and most sensitive $Z'$ boson decay channel in our model at colliders.\\

This work is organised as follows. Section~\ref{sect:model} describes the $B-L$ model under consideration. Section~\ref{sect:comput} illustrates the computational techniques adopted. The results are presented in sections~\ref{sect:Zp_disc} and \ref{sect:Zp_excl} for the $Z'$ boson sector. Finally, the conclusions are given in section~\ref{sect:conc}.

\section{The Model}\label{sect:model}
The model under study is the so-called ``pure'' or ``minimal''
$B-L$ model (see ref.~\cite{bbms} for conventions and references) 
since it has vanishing mixing between the $U(1)_{Y}$ 
and $U(1)_{B-L}$ gauge groups.
In the rest of this paper we refer to this model simply as the ``$B-L$
model''.  {In this model the} classical gauge invariant Lagrangian,
obeying the $SU(3)_C\times SU(2)_L\times U(1)_Y\times U(1)_{B-L}$
gauge symmetry, can be decomposed as:
\begin{equation}\label{L}
\mathscr{L}=\mathscr{L}_{YM} + \mathscr{L}_s + \mathscr{L}_f + \mathscr{L}_Y \, .
\end{equation}
The non-Abelian field strengths in $\mathscr{L}_{YM}$ are the same as in the SM
whereas the Abelian
ones can be written as follows:
\begin{equation}\label{La}
\mathscr{L}^{\rm Abel}_{YM} = 
-\frac{1}{4}F^{\mu\nu}F_{\mu\nu}-\frac{1}{4}F^{\prime\mu\nu}F^\prime _{\mu\nu}\, ,
\end{equation}
where
\begin{eqnarray}\label{new-fs3}
F_{\mu\nu}		&=&	\partial _{\mu}B_{\nu} - \partial _{\nu}B_{\mu} \, , \\ \label{new-fs4}
F^\prime_{\mu\nu}	&=&	\partial _{\mu}B^\prime_{\nu} - \partial _{\nu}B^\prime_{\mu} \, .
\end{eqnarray}
In this field basis, the covariant derivative is:
\begin{equation}\label{cov_der}
D_{\mu}\equiv \partial _{\mu} + ig_S T^{\alpha}G_{\mu}^{\phantom{o}\alpha} 
+ igT^aW_{\mu}^{\phantom{o}a} +ig_1YB_{\mu} +i(\widetilde{g}Y + g_1'Y_{B-L})B'_{\mu}\, .
\end{equation}
The ``pure'' or ``minimal'' $B-L$ model is defined by the condition $\widetilde{g} = 0$, that implies no mixing between the $B-L$ $Z'$ and SM $Z$ gauge bosons.

The fermionic Lagrangian (where $k$ is the
generation index) is given by
\begin{eqnarray} \nonumber
\mathscr{L}_f &=& \sum _{k=1}^3 \Big( i\overline {q_{kL}} \gamma _{\mu}D^{\mu} q_{kL} + i\overline {u_{kR}}
			\gamma _{\mu}D^{\mu} u_{kR} +i\overline {d_{kR}} \gamma _{\mu}D^{\mu} d_{kR} +\\
			  && + i\overline {l_{kL}} \gamma _{\mu}D^{\mu} l_{kL} + i\overline {e_{kR}}
			\gamma _{\mu}D^{\mu} e_{kR} +i\overline {\nu _{kR}} \gamma _{\mu}D^{\mu} \nu
			_{kR} \Big)  \, ,
\end{eqnarray}
 where the fields' charges are the usual SM and $B-L$ ones (in particular, $B-L = 1/3$ for quarks and $-1$ for leptons {with no distinction between generations, hence ensuring universality)}.
  The  $B-L$ charge assignments of the fields
  as well as the introduction of new
  fermionic  right-handed heavy neutrinos ($\nu_R$'s) and a
  scalar Higgs field ($\chi$, with charge $+2$ under $B-L$)  
  are designed to eliminate the triangular $B-L$  gauge anomalies and to ensure the gauge invariance of the theory, respectively.
  Therefore, a $B-L$  gauge extension of the SM gauge group
  broken at the TeV scale requires
  at least one new scalar field and three new fermionic fields which are
  charged with respect to the $B-L$ group.

The scalar Lagrangian is:
\begin{equation}\label{new-scalar_L}
\mathscr{L}_s=\left( D^{\mu} H\right) ^{\dagger} D_{\mu}H + 
\left( D^{\mu} \chi\right) ^{\dagger} D_{\mu}\chi - V(H,\chi ) \, ,
\end{equation}
{with the scalar potential given by}
\begin{equation}\label{new-potential}
V(H,\chi ) = - m^2H^{\dagger}H - \mu ^2\mid\chi\mid ^2 +
  \lambda _1 (H^{\dagger}H)^2 +\lambda _2 \mid\chi\mid ^4 + \lambda _3 H^{\dagger}H\mid\chi\mid ^2  \, ,
\end{equation}
{where $H$ and $\chi$ are the complex scalar Higgs 
doublet and singlet fields, respectively.}

Finally, the Yukawa interactions are:
\begin{eqnarray}\nonumber
\mathscr{L}_Y &=& -y^d_{jk}\overline {q_{jL}} d_{kR}H 
                 -y^u_{jk}\overline {q_{jL}} u_{kR}\widetilde H 
		 -y^e_{jk}\overline {l_{jL}} e_{kR}H \\ \label{L_Yukawa}
	      & & -y^{\nu}_{jk}\overline {l_{jL}} \nu _{kR}\widetilde H 
	         -y^M_{jk}\overline {(\nu _R)^c_j} \nu _{kR}\chi +  {\rm 
h.c.}  \, ,
\end{eqnarray}
{where $\widetilde H=i\sigma^2 H^*$ and  $i,j,k$ take the values $1$ to $3$},
where the last term is the Majorana contribution and the others the usual Dirac ones.

Neutrino mass eigenstates, obtained after applying the see-saw mechanism, will be called $\nu_l$ and $\nu_h$, where the first are the SM-like ones. With a reasonable choice of Yukawa couplings, the heavy neutrinos can have masses $m_{\nu_h} \sim \mathcal{O}(100)$ GeV $\ll M_{Z'_{B-L}}$. In such a case, the decay of the $Z'_{B-L}$ gauge boson into pairs of heavy neutrinos is allowed, therefore modifying quantitatively all the other decay channels. The corresponding Branching Ratio (BR) depends upon both heavy neutrino and $Z'_{B-L}$ masses and can reach  up to $\sim 18\%$, while $BR(Z'\rightarrow \ell^+ \ell^-)$ varies between $12.5\%$ and $15.5\%$ ($\ell =e,\mu$).
For a more exhaustive explanation, see ref.~\cite{bbms}. To be definite, the following analysis has been done for degenerate heavy neutrino masses with $m_{\nu ^1_h}=m_{\nu ^2_h}=m_{\nu ^3_h}=200$ GeV, value that can lead to an interesting phenomenology \cite{bbms}\footnote{Although they do not modify the $Z'$ boson properties significantly, for completeness we state here also the chosen scalar masses and mixing angle: $m_{h_1}=125$ GeV, $m_{h_2}=450$ GeV and $\alpha=0.01$. Such values are allowed by the study of the unitarity bound \cite{Basso:2010jt}, as well as of the triviality bound \cite{Basso:2010jm}, of the scalar sector.}. 

An important feature of the $Z'$ gauge boson in the $B-L$ model is the chiral structure of its couplings to fermions: since the $B-L$ charges do not distinguish between left-handed and right-handed fermions, the $B-L$ neutral current is purely vector-like, with a vanishing axial part\footnote{That is, $\displaystyle g_{Z'}^V = \frac{g_{Z'}^L+g_{Z'}^R}{2}$, $\displaystyle g_{Z'}^A = \frac{g_{Z'}^R-g_{Z'}^L}{2}=0$, hence $g_{Z'}^R=g_{Z'}^L$. On the contrary, Majorana neutrinos have pure axial couplings \cite{Perez:2009mu}.}. As a consequence, we decided not to study the asymmetries of the decay products stemming from $Z'_{B-L}$ boson, given their trivial distribution at the peak, the region we study here. However, asymmetries become important in the interference region, especially just before the $Z'$ boson peak, where the $Z-Z'$ interference will effectively provide an asymmetric distribution somewhat milder than the case in which there is no $Z'$ boson. This is a  powerful method of discovery and identification of a $Z'$ boson and it will be reported on separately. 

\section{Computational details}\label{sect:comput}

The study we present in this paper has been performed using the CalcHEP package \cite{calchep}. The model 
under discussion had been implemented in this package using the 
LanHEP tool \cite{lanhep}, as already discussed in Ref.~\cite{bbms}. 

The process we are interested in is di-lepton production. We define our signal as $pp\rightarrow \gamma,\,Z,\,Z'_{B-L}\rightarrow  \ell^+ \ell^-$ ($\ell=e,\,\mu$), i.e., all possible sources together with their mutual interferences, and the background as $pp\rightarrow \gamma,\,Z\rightarrow \ell^+ \ell^-$ ($\ell=e,\,\mu$), i.e., SM Drell-Yan production (including interference). No other sources of background, such as $WW$, $ZZ$, $WZ$, or $t\overline{t}$ have been taken into account. These can be suppressed or/and are insignificant \cite{Ball:2007zza,Abazov:2010ti}.
For both the signal and the background, we have assumed standard
acceptance cuts (for both electrons and muons) at the LHC:
\begin{equation}\label{LHC_cut}
p_T^\ell > 10~{\rm GeV},\qquad |\eta^\ell|<2.5\qquad (\ell=e,\,\mu),
\end{equation} 
and we apply 
the following requirements on the di-lepton invariant mass,  $M_{\ell \ell}$, depending on whether we are considering electrons or muons.
We distinguish two different scenarios: an ``early'' one (for $\sqrt{s}=7$, $10$ TeV) and an ``improved'' one  (for $\sqrt{s}=14$ TeV), and, in computing the signal significances, we will select a window as large as either one width of the $Z'_{B-L}$ boson or twice the di-lepton mass resolution\footnote{We take  the CMS di-electron and di-muon mass resolutions \cite{CMSdet,Ball:2007zza} as a typical LHC environment. ATLAS resolutions \cite{ATLASdet} do not differ substantially.}, 
whichever the largest. The half windows in the invariant mass distributions respectively read, for the ``early scenario'':
\begin{eqnarray}\label{LHC_ris_el}
\mbox{electrons: }\; |M_{ee}-M_{Z'}| &<& 
\mbox{max} \left( \frac{\Gamma_{Z'}}{2},\; \left( 0.02\frac{M_{Z'}}{\rm GeV} \right) {\rm GeV}\; \right),\\ \label{LHC_ris_mu}
\mbox{muons: }\; |M_{\mu\mu}-M_{Z'}| &<& 
\mbox{max} \left( \frac{\Gamma_{Z'}}{2},\; \left( 0.08\frac{M_{Z'}}{\rm GeV} \right) {\rm GeV}\; \right),
\end{eqnarray}
and for the ``improved scenario'':
\begin{eqnarray}\label{LHC_ris_el_imp}
\mbox{electrons: }\; |M_{ee}-M_{Z'}| &<& 
\mbox{max} \left( \frac{\Gamma_{Z'}}{2},\; \left( 0.005\frac{M_{Z'}}{\rm GeV} \right) {\rm GeV}\; \right),\\ \label{LHC_ris_mu_imp}
\mbox{muons: }\; |M_{\mu\mu}-M_{Z'}| &<& 
\mbox{max} \left( \frac{\Gamma_{Z'}}{2},\; \left( 0.04\frac{M_{Z'}}{\rm GeV} \right) {\rm GeV}\; \right).
\end{eqnarray}
Our choice reflects the fact that what we will observe is in fact the convolution 
between the Gaussian detector resolution and the Breit-Wigner 
shape of the peak, and such convolution will be dominated by the largest of the two. Our approach is to take the convolution width exactly equal to the resolution width or to the peak width, whichever is largest\footnote{In details, for resolutions below $\Gamma /2$, we take the convolution equal to the resolution width. For resolutions above $3\Gamma$, we take the convolution equal to the peak width. When the resolution $\in \left[ \Gamma /2,3\Gamma\right]$, the convolution is taken as a linear interpolation between the two regimes.}, and to count all the events within this window. Finally, only $68\%$ of signal events are considered: intrinsically, when the the peak width is dominating, effectively (by rescaling the signal), otherwise.

\begin{figure}[!h]
  \includegraphics[angle=0,width=0.8\textwidth ]{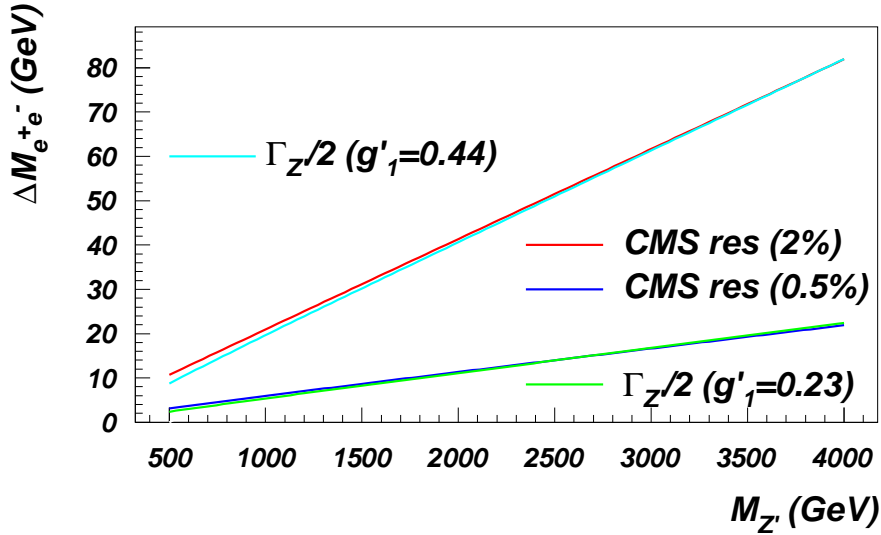}
  \caption{\it Comparison of the CMS electron resolution according to eqs.~(\ref{LHC_ris_el}) and (\ref{LHC_ris_el_imp}), with the two different constant terms as described in the text, compared to $\, \Gamma _{Z'}/2$ for $g'_1=0.23$ and $g'_1=0.44$.}
  \label{CMS_res}
\end{figure}

In figure \ref{CMS_res} we compare the LHC resolutions for electrons for the two aforementioned scenarios (eqs.~(\ref{LHC_ris_el}) and (\ref{LHC_ris_el_imp})) with $\Gamma_{Z'}/2$.
It is clear that, whichever the $Z'_{B-L}$ mass, for a value of the coupling $g'_1$ smaller than roughly $0.4$, the peak will be dominated by the early experimental resolution,
%
%
%
i.e., the half window 
will contain an amount of signal as big as the one produced with
$|M_{\ell \ell}-M_{Z'}| = \Gamma_{Z'}/2$. The region of interest in the parameter space we are going to study almost always fulfils the condition $g'_1<0.4$, as we will
see from the plots in the following section. The muon resolution is much worse and in such a plot it would be an order of magnitude higher than the other curves. Hence, for this final state, the peak is always dominated by the experimental resolution, for the values of the gauge coupling we are considering. Moreover, the better resolution for the electron channel means that the sensitivity of this channel will always be better than (or equal to) the muon channel.

In the next section we will compare the LHC and Tevatron discovery reach. In the derivation of the experimental constraints, we refer to the latest publications, being the D$\O$ analysis of Ref.~\cite{Abazov:2010ti} for the electron case and the CDF analysis of Ref.~\cite{Aaltonen:2011gp} for the muon final state. Hence, we have considered the typical acceptance cuts (for  electrons and muons) for the respective detector:
\begin{eqnarray}\label{Tev_cut}
p_T^e > 25~{\rm GeV}, &\qquad& |\eta^e|<1.1, \\
p_T^\mu > 18~{\rm GeV}, &\qquad& |\eta^\mu|<1,
\end{eqnarray}
and the following requirements on the di-lepton invariant mass,  $M_{\ell \ell}$, depending on whether we are considering electrons or 
muons\footnote{We take the D$\O$ di-electron \cite{D0res} and the CDF di-muon \cite{Aaltonen:2011gp} mass resolutions as a typical Tevatron environment, in accordance with the most up-to-date limits.}:
\begin{eqnarray}\label{Tev_ris}
\mbox{electron: }\; |M_{ee}-M_{Z'}| &<& 
\mbox{max} \left( \frac{\Gamma_{Z'}}{2},\; \left( 0.16 \sqrt{\frac{M_{Z'}}{\rm GeV}}{\rm GeV}+ 0.04\frac{M_{Z'}}{\rm GeV} \right) {\rm GeV}\; \right),\\
\mbox{muons: }\; |M_{\mu\mu}-M_{Z'}| &<& 
\mbox{max} \left( \frac{\Gamma_{Z'}}{2},\; \left( 0.017\%\, \left(\frac{M_{Z'}}{\rm GeV}\right) ^2 \right) {\rm GeV}\; \right).
\end{eqnarray}

The selection of an invariant mass window centred at the $Z'$ boson mass is comparable to the standard experimental analysis, as in Ref.~\cite{Abazov:2010ti} (the electron channel at D$\O$), where signal and background are integrated from $M_{Z'}-10\Gamma _{Z'}$ (where $\Gamma _{Z'}$ is the $Z'$ boson width obtained by rescaling the SM-$Z$ boson width by the ratio of the $Z'$ to the $Z$ boson mass) to infinity.
Since the background (in proximity of the narrow resonance) can be reasonably thought of as flat, while the signal is not, the procedure we propose enhances the signal more than the background and it is expected to be more sensitive. Reference \cite{Aaltonen:2011gp} applies a different strategy and figure~\ref{contour7_ecal} shows that our procedure is comparable,
 although less involved.
A Bayesian approach is being used at the LHC~\cite{Khachatryan:2010fa}, similar to the CDF case. Hence, we present our results for a comparison `a posteriori'.


In our analysis we use a definition of signal significance $\sigma$, as follows. 
In the region where the number of both signal ($s$) and background ($b$) events is ``large'' (here taken to be bigger than 20), we use a definition of significance based on Gaussian statistics:
\begin{equation}\label{signif}
{\sigma} \equiv {\it s}/{\sqrt{\it b}}.
\end{equation}
Otherwise, in case of smaller statistics, we used the Bityukov algorithm~\cite{Bityukov}, which basically uses the Poisson distribution instead of the approximate Gaussian one.

Finally, as in \cite{bbms,bbmp}, we used {\rm{CTEQ6L}} \cite{CTEQ} as default Parton Distribution Functions (PDFs), evaluated at the scale $Q^2=M_{\ell \ell}^2$.
%
%
The leading order (LO) cross sections are multiplied by a mass independent $k-$factor of $1.3$ \cite{Carena}, both for the cross sections evaluated at the Tevatron (as in Refs.~\cite{Abazov:2010ti,Aaltonen:2011gp,D0}) and at the LHC (as in Ref.~\cite{Khachatryan:2010fa}\footnote{Notice that in Ref.~\cite{Khachatryan:2010fa} the k-factor used was mass-dependent. Here we use the average value.}), to get in agreement with the Next-to-Next-to-Leading-Order (NNLO) QCD corrections.

Typical detector resolution has effectively been taken into account by our procedure, that consists in counting all the events that occur within the window (in invariant mass) previously described, and by rescaling to $68\%$ the signal events when the peak is dominated by the experimental resolution.
Nonetheless, our simulation does not account for Initial State Radiation (ISR) effects. 
ISR can have two main sources: QED-like ISR (i.e., photon emission), that has the effect of shifting the peak and of creating a tail towards smaller energy, and QCD-like ISR (i.e., gluon emission), that has similar effects and might also induce trigger issues
in the intent of removing backgrounds (e.g., by cutting on final state jets). Although we are aware of such effects, we believe that their analysis goes beyond the scope of this paper and it will be reported on separately.
Altogether, we are confident that, while particular aspects of
our analysis may be sensitive to such effects, the general picture will not depend upon these substantially (see section~\ref{validation}).
Also, the only background considered here was the irreducible SM Drell-Yan. Reducible backgrounds, photon-to-electron conversion, efficiencies in reconstructing electrons/muons, jets faking leptons etc., whose overall effect is to deplete the signal, were neglected (being $t\overline{t}$ the most important source, at the level of $ \lesssim 10\%$). However, for this analysis they are not quantitatively important \cite{Abulencia:2005ix,Ball:2007zza,Abazov:2010ti}.
The net effect of the factors above is usually regarded as an overall reduction of the total acceptance, being the lepton identification the most important source, about $80\div 90\%$ per each lepton. We comment on this in the conclusions.

\subsection{Validation}\label{validation}
We can now quantitatively compare our simulation to the literature. Ref.~\cite{D0} contains a comprehensive analysis of expected backgrounds for several $Z'$ boson masses\footnote{The simulation therein is modelled using the {\rm{PYTHIA}} \cite{Sjostrand:1993yb} Monte Carlo event generator, with {\rm{CTEQ6L1}} PDFs, and then processed through the standard D$\O$ detector simulation based on {\rm{GEANT3}} \cite{geant}.} at $3.6$ fb$^{-1}$ of integrated luminosity. Table~\ref{D0-comparison} shows the comparison of our expectations with theirs, based on table~II in \cite{D0}, where the compatibility is, as usual, defined as:
\begin{equation}
\frac{Ev - Ev_{D\O}}{\sigma _{D\O}}\, .
\end{equation}
\begin{table}[h]
\begin{center}
\begin{tabular}{|c|c|c|c||c|c|}
\hline
& & \multicolumn{2}{|c||}{D$\O$ analysis} & \multicolumn{2}{|c|}{Our simulation}\\
\hline \vspace{-0.25cm}
 $M_{Z'}$ & Mass Window      & Expected Bkg & Signal & Expected Bkg & Compatibility  \\ 
 (GeV)    & Lower Limit (GeV)& Events   & Acceptance & Events   & Level \\
\hline
400 & 353  & 22.4 $\pm$ 0.7	& 0.172  & 23.2  & 1.1 \\  
500 & 445  & 7.92 $\pm$ 0.22	& 0.188  & 8.23  & 1.4 \\ 
600 & 536  & 2.93 $\pm$ 0.07	& 0.199  & 3.00  & 1.0 \\  
700 & 626  & 1.052 $\pm$ 0.025	& 0.207  & 1.110 & 2.3 \\ 
750 & 673  & 0.631 $\pm$ 0.016	& 0.209  & 0.653 & 1.4 \\ 
800 & 718  & 0.384 $\pm$ 0.010	& 0.211  & 0.391 & 0.7 \\ 
850 & 762  & 0.222 $\pm$ 0.006	& 0.212  & 0.235 & 2.2 \\
900 & 810  & 0.134 $\pm$ 0.004	& 0.216  & 0.135 & 0.4 \\ 
950 & 858  & 0.0701 $\pm$ 0.0023& 0.214  & 0.0750& 2.1 \\ 
1000& 902  & 0.0410 $\pm$ 0.0015& 0.216  & 0.0440& 2.0 \\
\hline
\end{tabular}
\end{center}
\vskip -0.5cm
\caption{\it Comparison of our simulation to the D$\O$ analysis. The left hand side of the table is taken from Table II of Ref.~\cite{D0}.
\label{D0-comparison}}
\end{table}

It is clear that our simulation is reasonable despite the lack of detector simulation, since it reproduces the D$\O$ backgrounds within two standard deviations. Nonetheless, the limits extracted from these data are quite looser compared to those we will derive in the next section, therefore we will not show them.

\section{$Z'$ Boson Sector: Discovery Power}\label{sect:Zp_disc}
In this section we determine the discovery potential of the LHC considering several center-of-mass (CM) energies, $7$, $10$ and $14$ TeV, using the expected integrated luminosities. In the following subsection we present the latest available experimental constraints and we compare our results for the LHC to the expected ultimate reach at the Tevatron, for $\sim 10$ fb$^{-1}$~\cite{Tevatron_reach}.

\begin{figure}[!h]
  \includegraphics[angle=0,width=0.8\textwidth ]{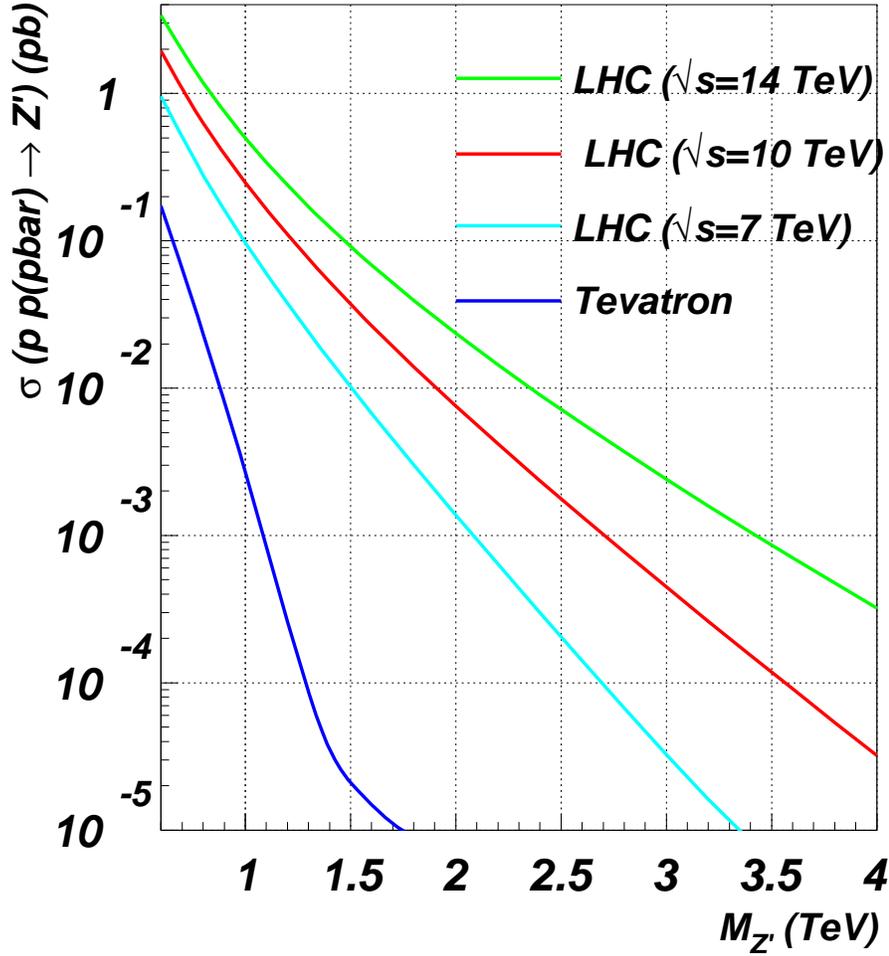}
  \caption{\it Cross sections for $pp(\overline{p}) \rightarrow Z'_{B-L}$ at the Tevatron and at the LHC (for $\sqrt{s}=7,10$ and $14$ TeV) for $g'_1=0.1$.}
  \label{Zp_xs}
\end{figure}

The production cross sections for the process $pp(\overline{p}) \rightarrow Z'_{B-L}$ for $g'_1=0.1$ are shown in figure~\ref{Zp_xs}.
Note that although at the Tevatron the production cross section is 
smaller than at the LHC, the integrated luminosities we are 
considering here for the LHC at 7 and 10 TeV (i.e., $1$ fb$^{-1}$) are smaller than for the Tevatron (i.e., $10$ fb$^{-1}$).

\subsection{LHC at $\boldsymbol{\sqrt{s}=7}$ TeV}
\begin{figure}[!h]
  \subfloat[]{ 
  \label{contour7_ecal}
  \includegraphics[angle=0,width=0.48\textwidth ]{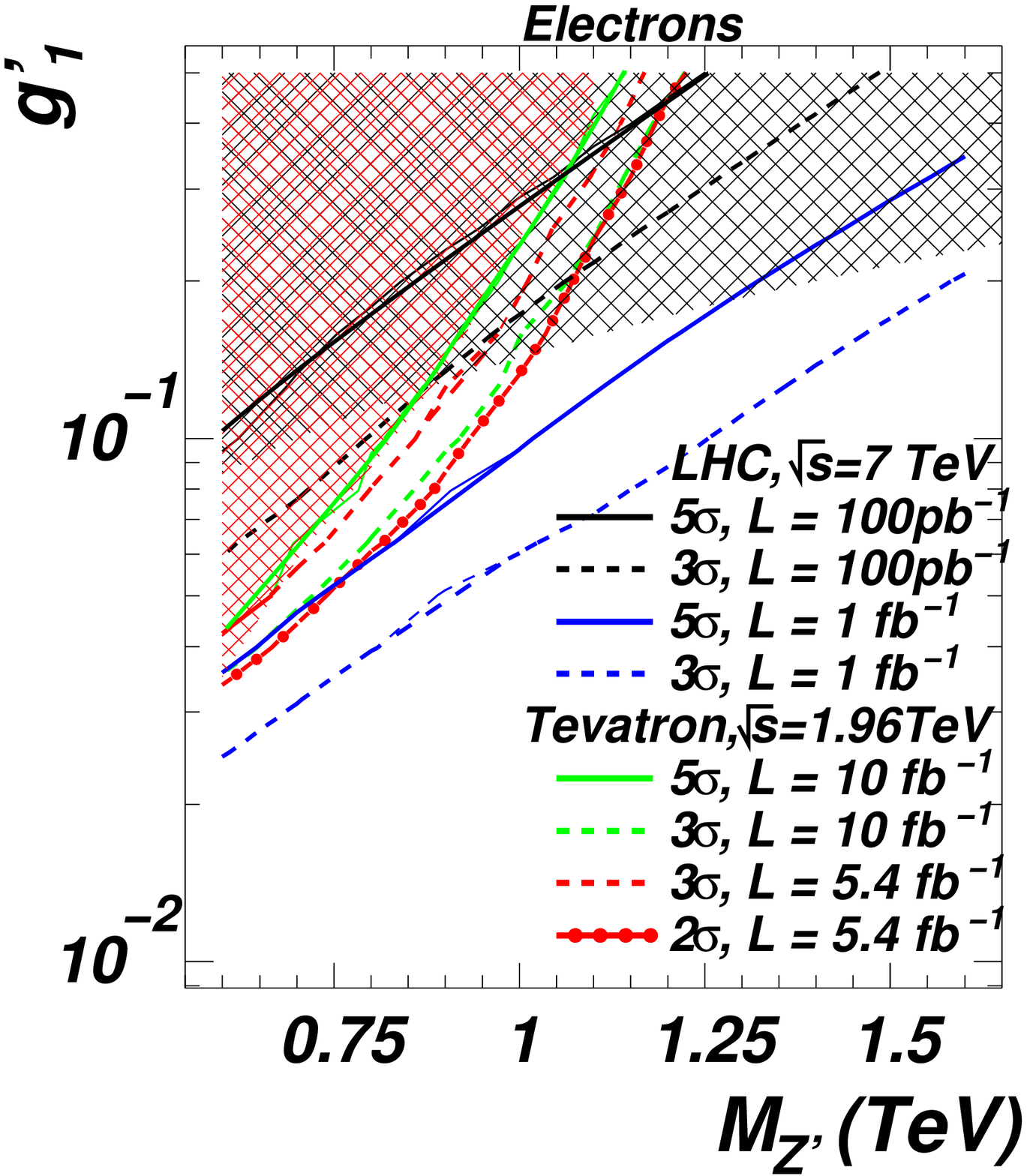}}
  \subfloat[]{
  \label{contour7_pt}
  \includegraphics[angle=0,width=0.48\textwidth ]{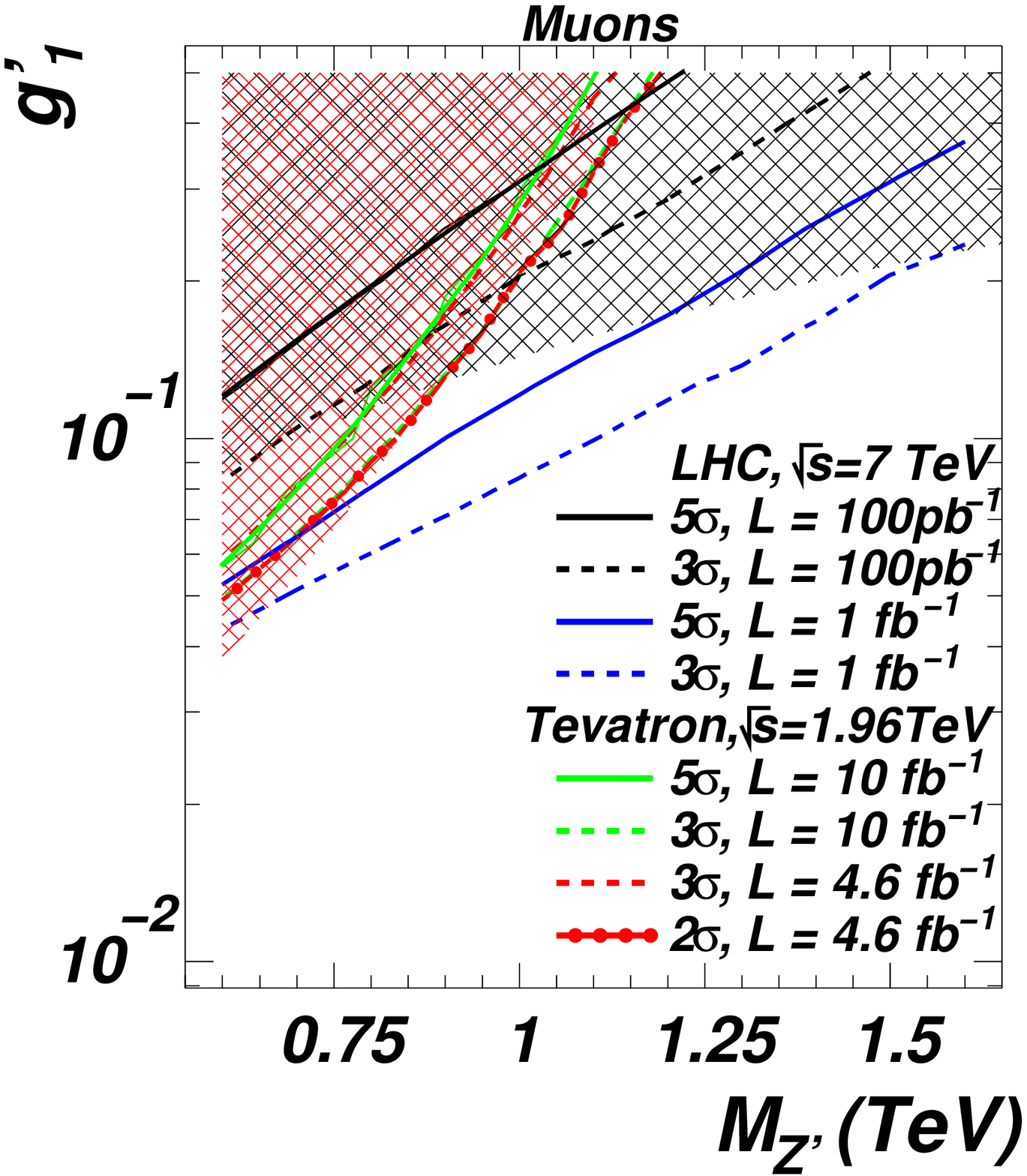}}
  \vspace*{-0.5cm}
  \caption{\it Significance contour levels plotted against $g_1'$
and $M_{Z'}$ at the LHC for $\sqrt{s}=7$ TeV for $0.1-1$ fb$^{-1}$ and at the Tevatron ($\sqrt{s}=1.96$ TeV) for (\ref{contour7_ecal}, electrons) $5.4-10\,\mbox{fb}^{-1}$ and (\ref{contour7_pt}, muons) $4.6-10\,\mbox{fb}^{-1}$ of integrated luminosity. The shaded areas correspond to the region of parameter space excluded
experimentally, in accordance with eq.~(\ref{LEP_bound}) (LEP bounds, in black) and table~\ref{mzp-low_bound} (Tevatron bounds, in red).}
  \label{contour7}
\end{figure}

The first years of the LHC work will be at a CM energy of $7$ TeV, where the total integrated luminosity is likely to be of the order of $1$ fb$^{-1}$. Figure~\ref{contour7} shows the discovery potential under these conditions, as well as the most recent limit from LEP \cite{Cacciapaglia:2006pk}:
\begin{equation}\label{LEP_bound}
\frac{M_{Z'}}{g'_1} \geq 7\; \rm{TeV}\, 
\end{equation}
and from the Tevatron\footnote{Notice that these are the most conservative limits, as they are evaluated for decoupled heavy neutrinos, i.e., with masses bigger than $M_{Z'}/2$.} (from the D$\O$ analysis of Ref.~\cite{Abazov:2010ti} using $5.4\,\mbox{fb}^{-1}$ and the CDF analysis of Ref.~\cite{Aaltonen:2011gp} using $4.6\,\mbox{fb}^{-1}$ of data, respectively for electrons and muons in the final state). The Tevatron limits for the $Z'_{B-L}$ boson are shown in table~\ref{mzp-low_bound} (for selected masses and couplings).
\begin{table}[h]
\begin{center}
\begin{tabular}{|c|c||c|c|}
\hline
\multicolumn{2}{|c||}{$p\overline{p}\rightarrow e^+ e^-$} & \multicolumn{2}{|c|}{$p\overline{p}\rightarrow \mu^+ \mu^-$} \\
\hline
 $g_1'$         & $M_{Z'}$ (GeV) & $g_1'$         & $M_{Z'}$ (GeV)\\
\hline
0.0197 & 300  & 0.0179	& 300        \\  
0.0193 & 400  & 0.0189	& 400        \\  
0.0281 & 500  & 0.0456	& 500        \\ 
0.0351 & 600  & 0.0380	& 600        \\  
0.0587 & 700  & 0.0544	& 700        \\ 
0.0880 & 800  & 0.0830	& 800        \\ 
0.1350 & 900  & 0.1360	& 900        \\ 
0.2411 & 1000 & 0.2220	& 1000        \\
0.3880 & 1100 & 0.3380	& 1100        \\
\hline
\end{tabular}
\end{center}
\vskip -0.5cm
\caption{\it Lower bounds on the $Z'$ boson mass for selected $g_1'$ values in the $B-L$ model, at $95\%$ C.L.,
by comparing the collected data of Refs.~\cite{Abazov:2010ti,Aaltonen:2011gp} with our theoretical prediction for $p\overline{p}\rightarrow Z'_{B-L} \rightarrow e^+e^-(\mu ^+\mu ^-)$ at the Tevatron. 
\label{mzp-low_bound}}
\end{table}
In the same figure we also include for comparison the Tevatron discovery potential at the integrated luminosities used for the latest published analyses \cite{Abazov:2010ti,Aaltonen:2011gp} ($5.4\,\mbox{fb}^{-1}$ and $4.6\,\mbox{fb}^{-1}$ for electrons and muons, respectively) as well as the expected reaches at $\mathcal{L}=10\,\mbox{fb}^{-1}$.

Notice that the Tevatron excluded area are based on the actual data, while the dot-dashed $2\sigma$ curves are our theoretical curves. Thus, if from the one side theory cannot reproduce experiments, from the other side we are comparing two methods of extracting the results.
%
%
%
%
As mentioned previously, figure~\ref{contour7} shows that the procedures used in experimental analyses for the electron channel \cite{D0,Abazov:2010ti} are not quite optimised for maximizing the signal significance. The alternative analysis described in this work has the potential to improve sensitivities and can be easily developed even further.

It is then clear that the Tevatron will still be 
competitive with the LHC (for $\sqrt{s}=7$ TeV CM energy), especially in the lower mass region where the LHC requires $1$ fb$^{-1}$ to be sensitive to the same couplings as the Tevatron.
The LHC will be able to probe, at $5\sigma$ level, the $Z'_{B-L}$ boson for values of the coupling
down to $3.7-5.2 \cdot 10^{-2}$ (for electrons and muons respectively), while 
the Tevatron can be sensitive down to $4.2 \cdot 10^{-2}$ with electrons. 
The kinematical reach of the two machines is different. The LHC for $1$ fb$^{-1}$ can discover the $Z'_{B-L}$ boson up to masses of $1.20-1.25$ TeV, while at the Tevatron a $3\sigma$ evidence will be possible up to a value of the mass of $1$ TeV in the electron channel, for a suitable choice of the coupling. 
As is clear from figure~\ref{contour7_pt}, the muon channel at the Tevatron requires more than $10$ fb$^{-1}$ to start probing (at $3\sigma$) points in the $M_{Z'}-g'_1$ plane allowed by the CDF constraints, as in table~\ref{mzp-low_bound}. This total integrate luminosity appears to be more than what can be collected, due to the announced shutdown by the end of the year $2011$ \cite{Tevatron_reach,Tevatron_shutdown}.


\begin{figure}[!h]
  \subfloat[]{ 
  \label{contour7_ecal_LHC}
  \includegraphics[angle=0,width=0.48\textwidth ]{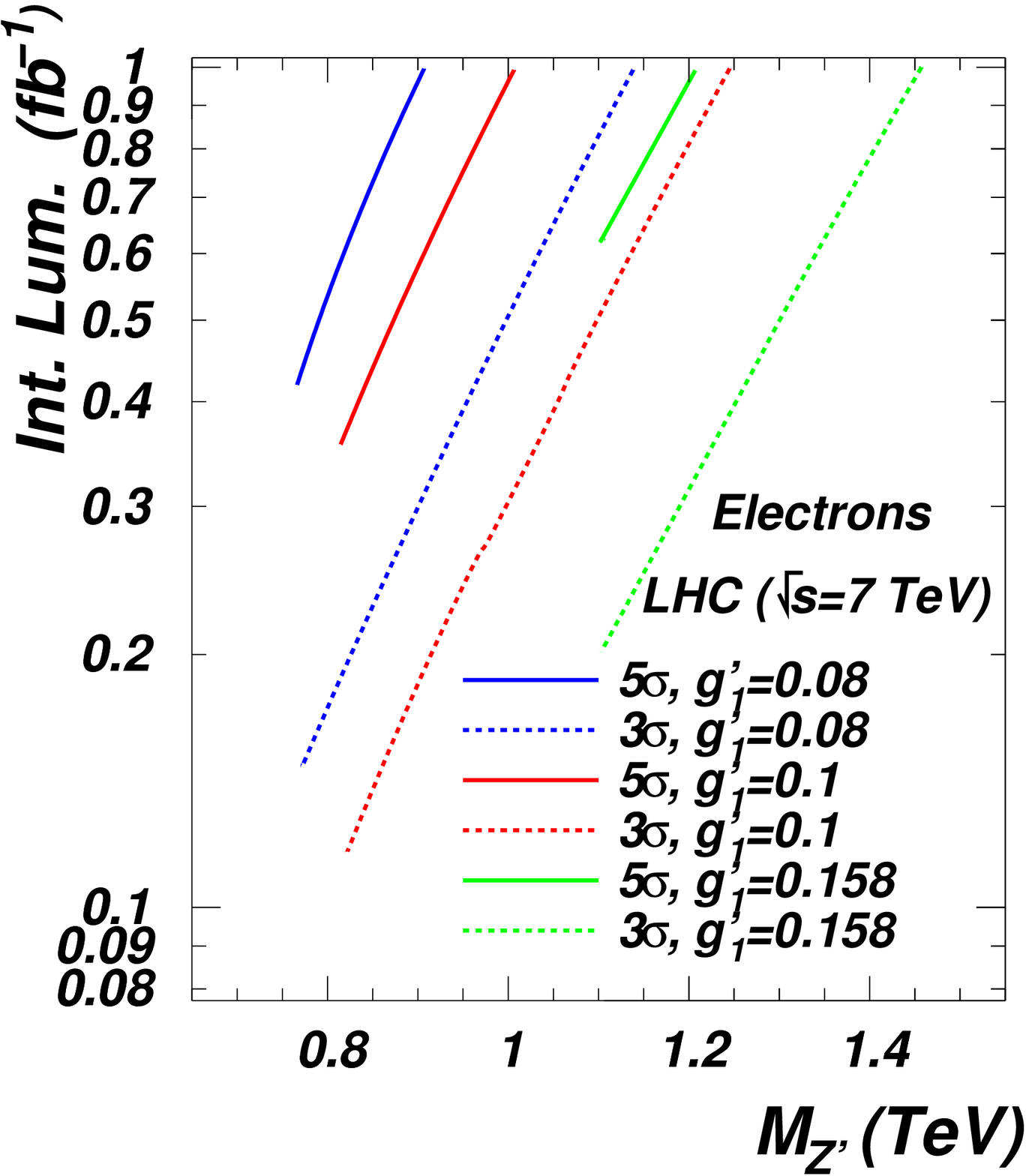}}
  \subfloat[]{
  \label{contour7_ecal_Tev}
  \includegraphics[angle=0,width=0.48\textwidth ]{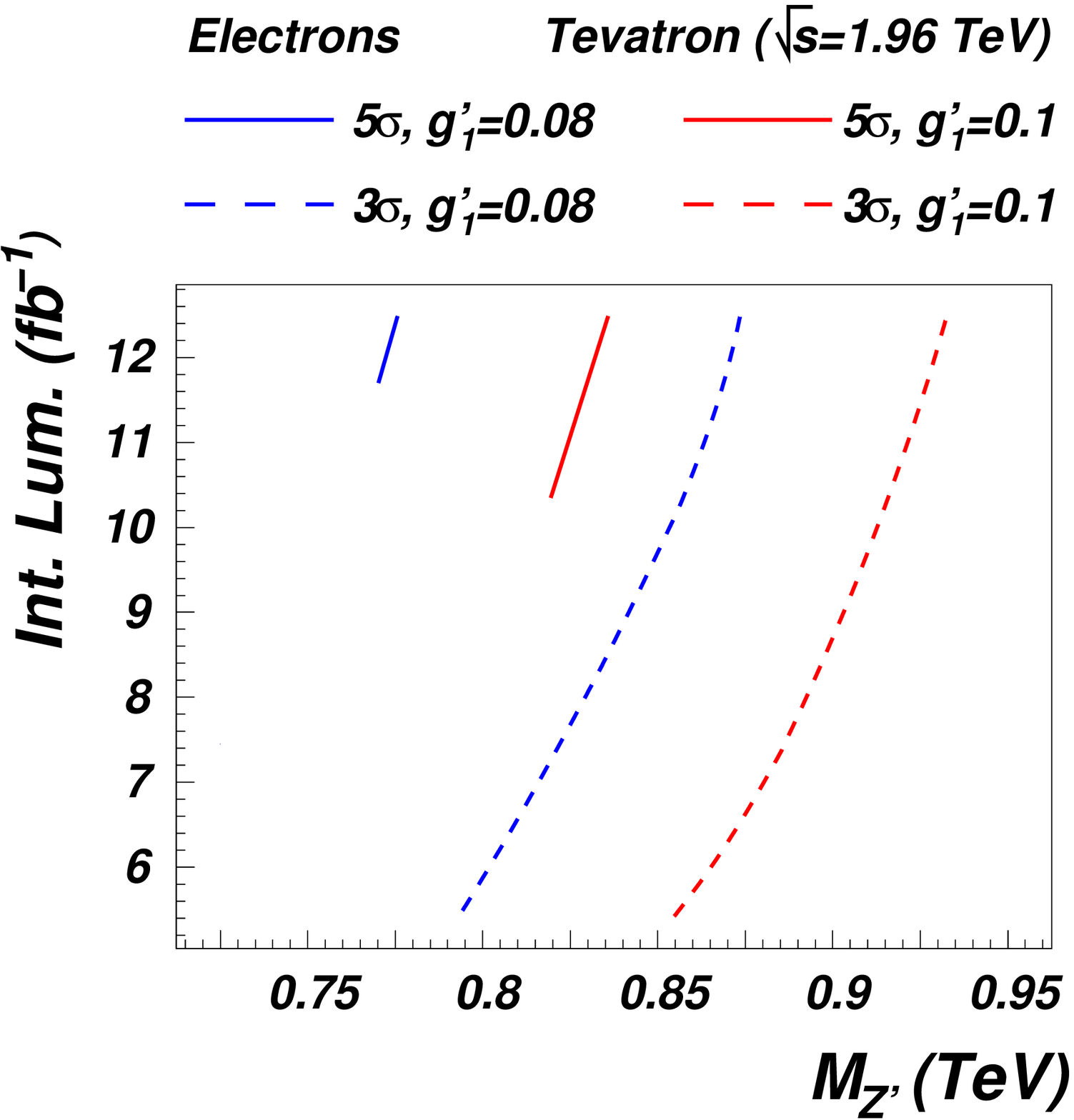}}\\
\flushleft{
  \subfloat[]{
  \label{contour7_pt_LHC}
  \includegraphics[angle=0,width=0.48\textwidth ]{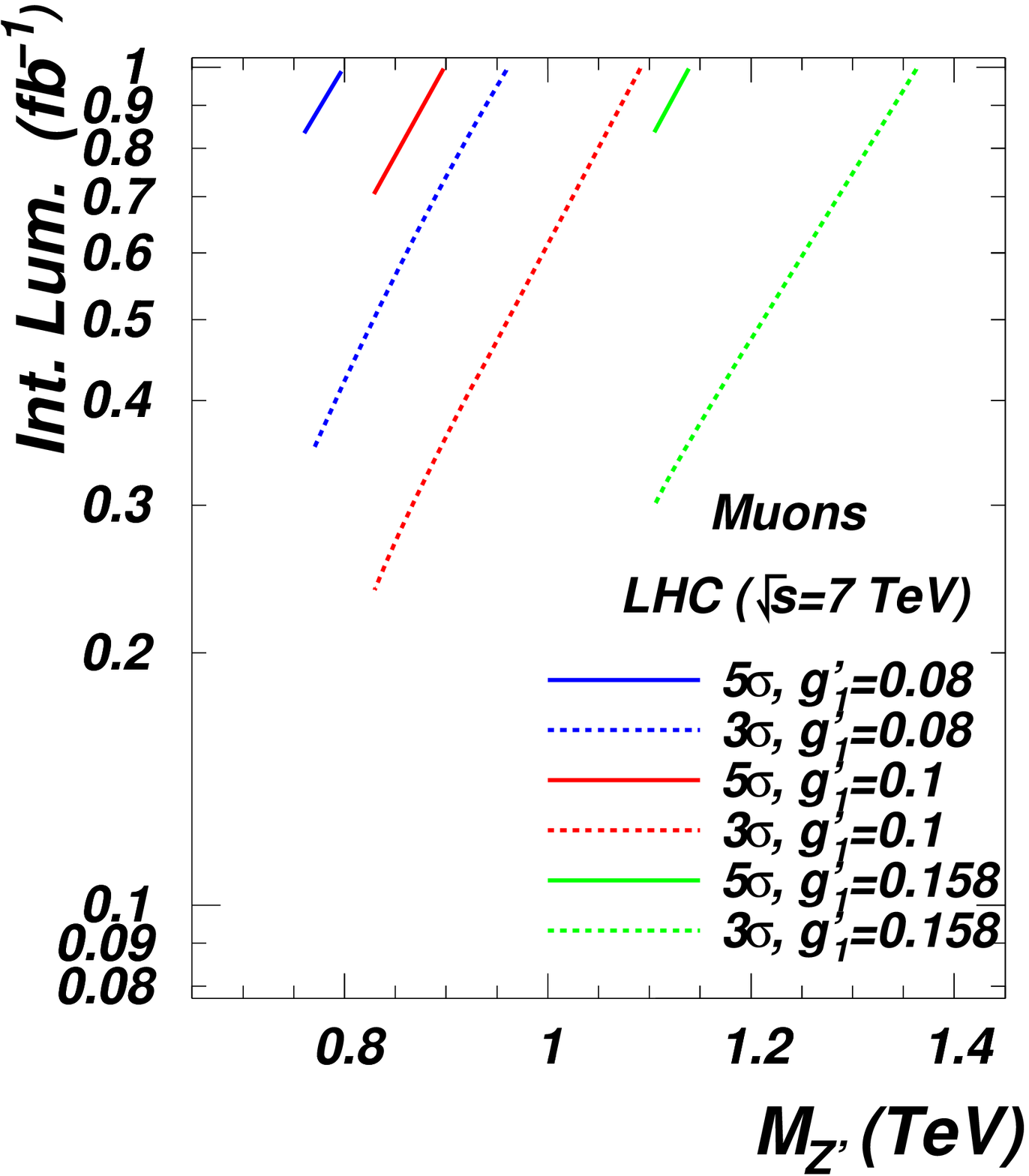}} } 
  \vspace*{-0.5cm}
  \caption{\it Integrated luminosity required for observation at $3\sigma$ and $5\sigma$ vs. $M_{Z'}$ for selected values of $g_1'$ at the LHC for $\sqrt{s}=7$ TeV for (\ref{contour7_ecal_LHC}) electrons and (\ref{contour7_pt_LHC}) muons and at the Tevatron ($\sqrt{s}=1.96$ TeV) for (\ref{contour7_ecal_Tev}) electrons (muons require more than $10$ fb$^{-1}$ and is, hence, not shown).
Only allowed combinations of masses and couplings are shown.}
  \label{lumi_vs_mzp_7TeV}
\end{figure}

Figure~\ref{lumi_vs_mzp_7TeV} shows the integrated luminosity required for $3\sigma$ evidence and $5\sigma$ discovery as a function of the $Z'_{B-L}$ boson mass for selected values of the coupling for both electron and muon final states at the LHC and, and for the electron channel only at the Tevatron. The muon channel at Tevatron requires more than $10$ fb$^{-1}$ to start probing the $Z'_{B-L}$ boson at $3\sigma$, and, hence, we do not present it.

We now fix some values for the coupling ($g'_1 = 0.158,\, 0.1,\, 0.08$ for the LHC analysis, $g'_1=0.1,\, 0.08$ for the Tevatron) and we see what luminosity is required for discovery at each machine. 
For $g'_1=0.1$ the LHC requires $0.35-0.70$ fb$^{-1}$ to be sensitive at $5\sigma$, with electrons and muons, respectively, while Tevatron requires $10$ fb$^{-1}$ with electrons. For the same value of the coupling, the Tevatron can discover the $Z'_{B-L}$ boson up to $M_{Z'}=840$ GeV, with electrons in the final state, with $12$ fb$^{-1}$ of data. The LHC can extend the Tevatron reach up to $M_{Z'}=1.0(0.9)$ TeV for $g'_1=0.1$. Regarding $g'_1=0.08$, a discovery can be made chiefly with electrons, requiring $0.4(12)$ fb$^{-1}$, for masses up to $900(780)$ GeV at the LHC(Tevatron). For muons, the LHC requires $0.85$ fb$^{-1}$ for masses up to $800$ GeV. Both machines will be sensitive at $3\sigma$ with much less integrated luminosities, requiring roughly $0.12-0.15(0.25-0.35)$ fb$^{-1}$ to probe the $Z'_{B-L}$ at the LHC for electrons(muons) in the final state, for $g'_1=0.1-0.08$. At the Tevatron, $5.5$ fb$^{-1}$ are required for probing at $3\sigma$ both values of the coupling, for electrons only. Finally, bigger values of the coupling, such as $g'_1=0.158$, can be probed just at the LHC, that is sensitive at $3\sigma$ for masses up $1.45(1.35)$ TeV using electrons(muons) and at $5\sigma$ for masses up to $1.2(1.15)$ GeV, requiring $0.2-0.6(0.3-0.9)$ fb$^{-1}$ at least, at $3\sigma-5\sigma$ for electrons(muons).

The $5\sigma$ discovery potential for the LHC at $\sqrt{s}=7$ TeV and for the Tevatron are summarised in table~\ref{5sigma_at_7TeV}, for selected values of couplings and integrated luminosities.
\begin{table}[h]
\begin{center}
\begin{tabular}{|c||c|c|c||c|c|c|}
\hline
LHC & \multicolumn{3}{|c||}{$pp\rightarrow e^+ e^-$} & \multicolumn{3}{|c|}{$pp\rightarrow \mu^+ \mu^-$} \\
\hline
$\int \mathcal{L}$ (fb$^{-1}$)  & $g'_1=0.08$ & $g'_1=0.1$ & $g'_1=0.158$ & $g'_1=0.08$ & $g'_1=0.1$ & $g'_1=0.158$ \\
\hline
0.2 & 820(-)   & 925(-)     & 1100(-)	 & -(-)    & -(-)     & -(-)      \\ 
0.3 & 900(-)   & 1000(800)  & 1200(-)	 & -(-)    & 850(-)   & 1100(-)   \\  
0.5 & 1000(775)& 1100(875)  & 1300(-)    & 825(-)  & 950(-)   & 1200(-)   \\ 
1   & 1130(900)& 1250(1000) & 1450(1200) & 950(800)& 1080(900)& 1360(1130)\\ 
\hline
\hline
Tevatron & \multicolumn{3}{|c||}{$pp\rightarrow e^+ e^-$} & \multicolumn{3}{|c|}{$pp\rightarrow \mu^+ \mu^-$} \\
\hline
$\int \mathcal{L}$ (fb$^{-1}$)  & $g'_1=0.08$ & $g'_1=0.1$ & $g'_1=0.158$ & $g'_1=0.08$ & $g'_1=0.1$ & $g'_1=0.158$ \\
\hline
8  & 825(-)   & 895(-)& -(-)& -(-)    & -(-) & -(-)       \\  
10 & 850(-)   & 915(825)& -(-)& -(-)    & -(-) & -(-)       \\ 
12 & 870(775) & 930(830)& -(-)& -(-)    & -(-) & -(-)    \\ 
\hline
\end{tabular}
\end{center}
\vskip -0.5cm
\caption{\it Maximum $Z'_{B-L}$ boson masses (in GeV) for a $3\sigma(5\sigma)$ discovery for selected $g_1'$ and integrated luminosities in the $B-L$ model, both at the LHC (for $\sqrt{s}=7$ TeV) and at the Tevatron (for $\sqrt{s}=1.96$ TeV). No numbers are quoted for already excluded configurations.}
\label{5sigma_at_7TeV}
\end{table}

\subsection{LHC at $\boldsymbol{\sqrt{s}=10}$ TeV}
\begin{figure}[!h]
  \subfloat[]{ 
  \label{contour10_ecal}
  \includegraphics[angle=0,width=0.48\textwidth ]{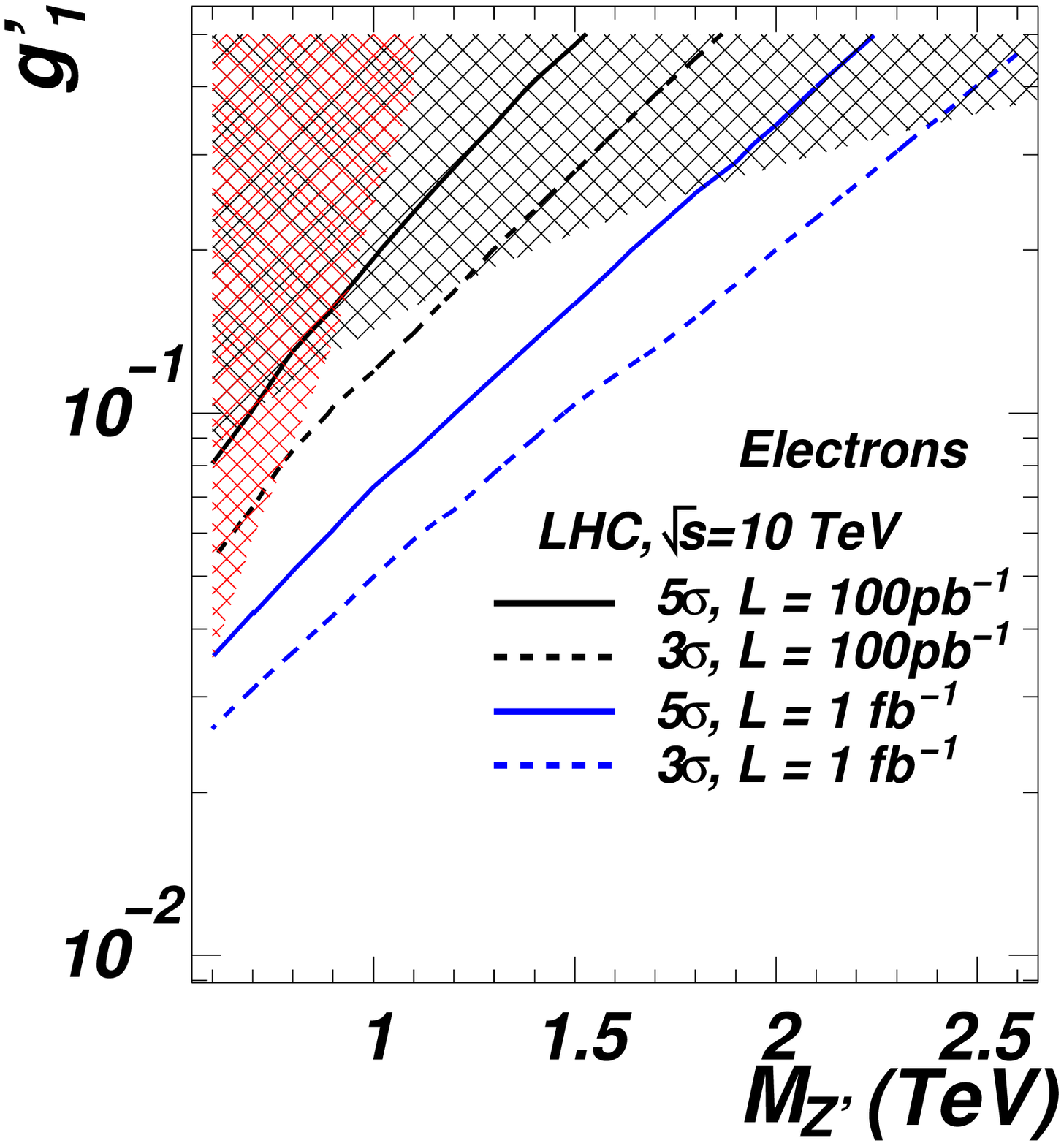}}
  \subfloat[]{
  \label{contour10_pt}
  \includegraphics[angle=0,width=0.48\textwidth ]{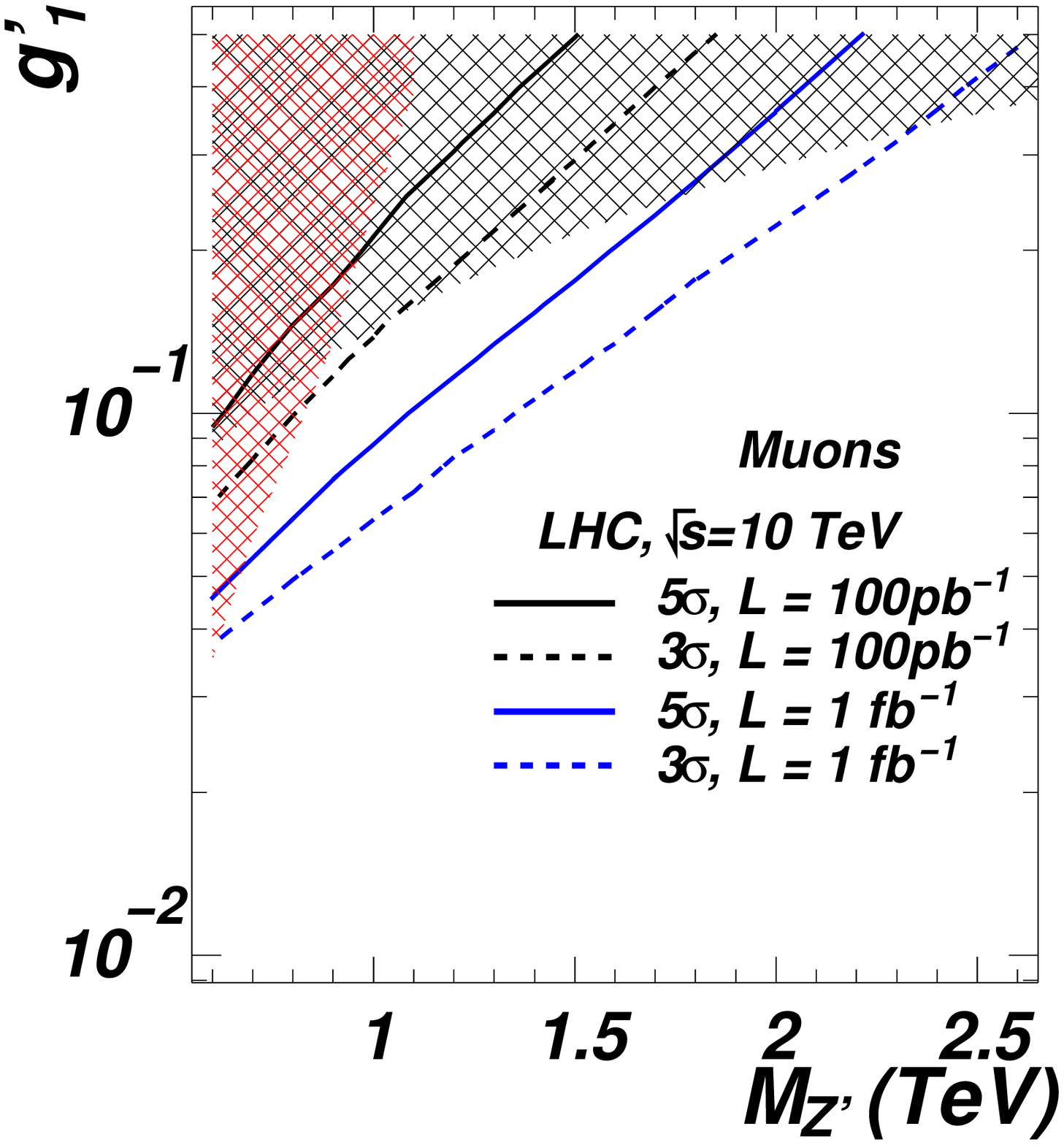}}
  \vspace*{-0.5cm}
  \caption{\it Significance contour levels plotted against $g_1'$
and $M_{Z'}$ at the LHC for $\sqrt{s}=10$ TeV for $100$ pb$^{-1}$ and $1$ fb$^{-1}$ for (\ref{contour10_ecal}) electrons and (\ref{contour10_pt}) muons. The shaded areas correspond to the region of parameter space excluded
experimentally, in accordance with eq.~(\ref{LEP_bound}) (LEP bounds, in black) and table~\ref{mzp-low_bound} (Tevatron bounds, in red).}
  \label{contour10}
\end{figure}

\begin{figure}[!h]
  \subfloat[]{ 
  \label{contour10_ecal_Lumi}
  \includegraphics[angle=0,width=0.48\textwidth ]{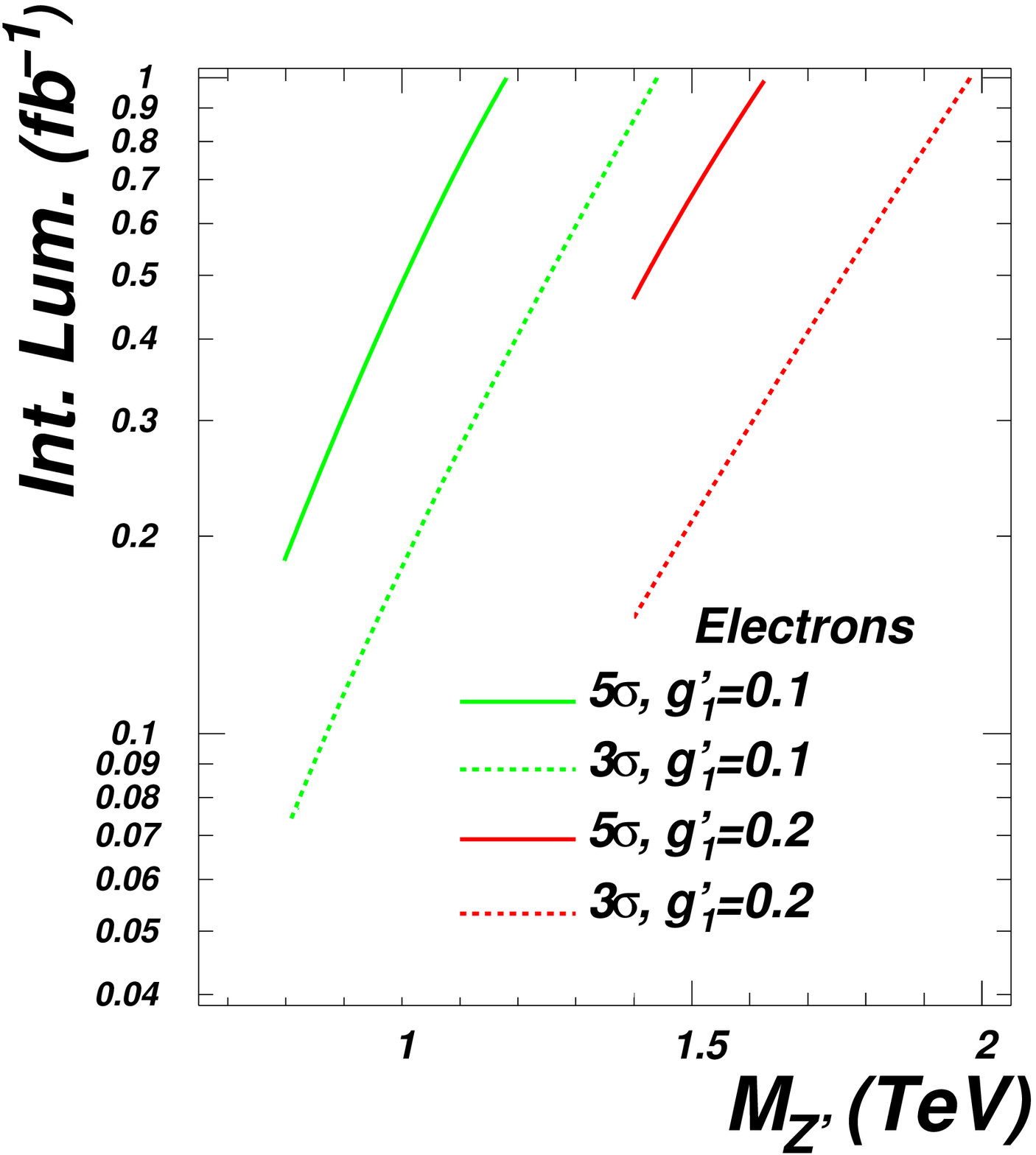}}
  \subfloat[]{
  \label{contour10_pt_Lumi}
  \includegraphics[angle=0,width=0.48\textwidth ]{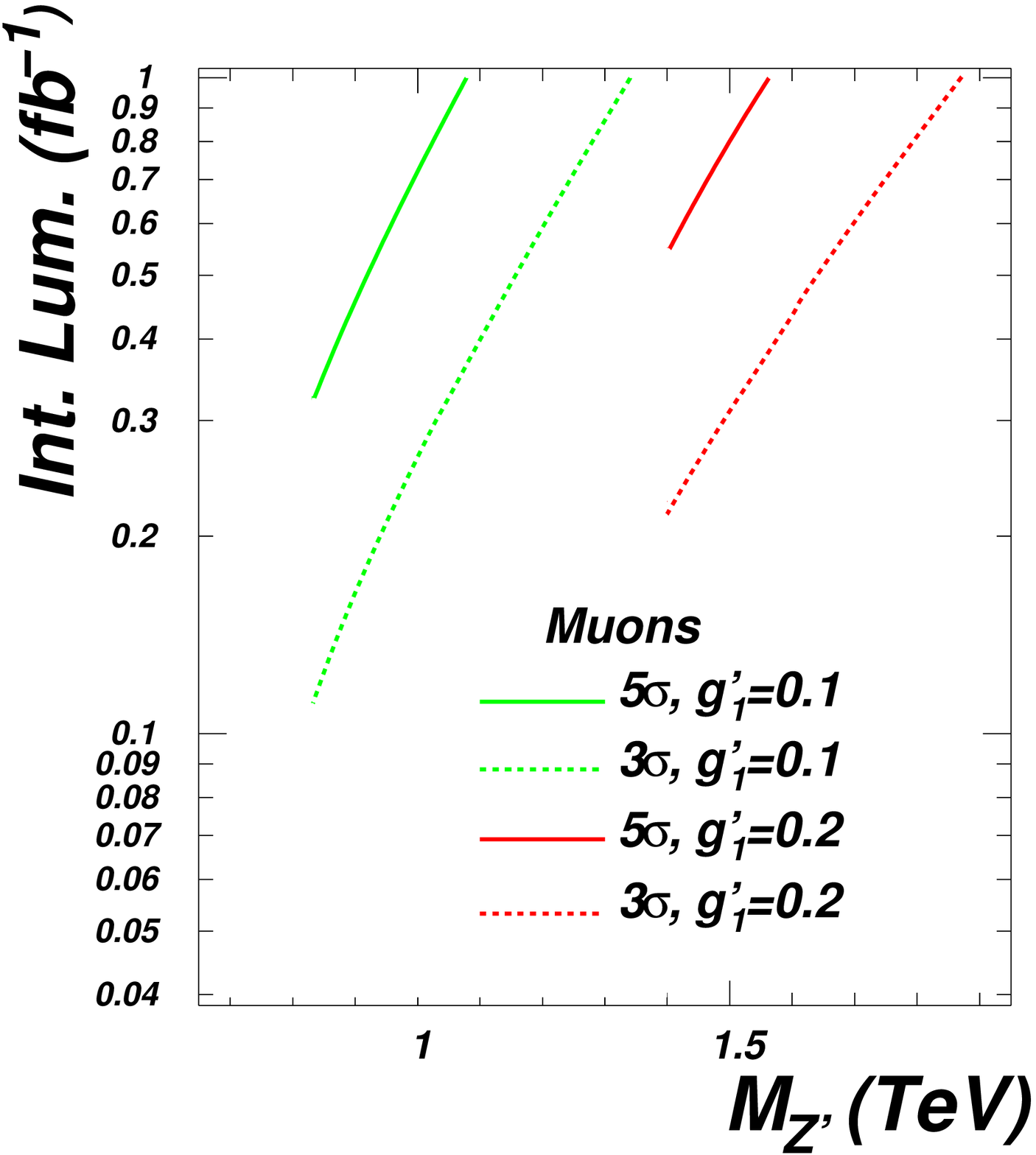}}
  \vspace*{-0.5cm}
  \caption{\it Integrated luminosity required for observation at $3\sigma$ and $5\sigma$ vs. $M_{Z'}$ for selected values of $g_1'$ at the LHC for $\sqrt{s}=10$ TeV for (\ref{contour10_ecal_Lumi}) electrons and (\ref{contour10_pt_Lumi}) muons. Only allowed combination of masses and couplings are shown.}
  \label{lumi_vs_mzp_10TeV}
\end{figure}

Figure~\ref{contour10} shows the discovery potential for the $Z'_{B-L}$ boson at the LHC for $10$ TeV CM energy, while figure~\ref{lumi_vs_mzp_10TeV} shows the integrated luminosity required for $3\sigma$ evidence as well as for $5\sigma$ discovery as a function of the $Z'_{B-L}$ boson mass, for selected values of the coupling constant, for both electron and muon channels.

%

If we compare figure~\ref{contour10} to figure~\ref{contour7}, we see that increasing the CM energy at the LHC allows to extend the kinematical reach of the machine towards bigger values of the mass for the $Z'_{B-L}$ boson. For $1$ fb$^{-1}$, the maximum observable masses for $\sqrt{s}=7$ TeV are $1.25(1.65)$ TeV and $1.20(1.50)$ TeV at $5(3)\sigma$, using electrons and muons, respectively. For $\sqrt{s}=10$ TeV the heaviest observable $Z'_{B-L}$ boson is for $M_{Z'}=1.8(2.4)$ TeV and $M_{Z'}=1.7(2.3)$ TeV, again at $5(3)\sigma$ and for electrons and muons, respectively, for a suitable choice of the coupling.

As we can see in figure~\ref{lumi_vs_mzp_10TeV}, $75(110)$ pb$^{-1}$ of data is the minimum integrated luminosity required to access the coupling $g'_1=0.1$ at $3\sigma$, in the electron(muon) final state, while $200(320)$ pb$^{-1}$ is required to access the same coupling at $5\sigma$, for a $Z'_{B-L}$ boson mass of $\sim 800$ GeV, where $Z'_{B-L}\rightarrow e^+ e^- (\mu ^+ \mu ^-)$. For a coupling of $0.2$, $150(500)$ pb$^{-1}$ and $210(550)$ pb$^{-1}$ are the minimum luminosities to start having evidences, at $3(5)\sigma$, of the $Z'_{B-L}$ boson decaying into electrons and muons, respectively.

If a $\sqrt{s}=10$ TeV run at the LHC is performed and is able to collect up to $1$ fb$^{-1}$ of data, it would be possible to discover(observe) a $Z'_{B-L}$ for masses up to $1.2(1.45)$ TeV and $1.10(1.35)$ TeV (at $5(3)\sigma$) by looking at electrons and muons in the final state, respectively, for $g'_1=0.1$, and $1.60(1.95)$ TeV and $1.55(1.85)$ for $g'_1=0.2$.

The $5\sigma$ discovery potential for the LHC at $\sqrt{s}=10$ TeV is summarised in table~\ref{5sigma_at_10TeV}, for selected values of couplings and integrated luminosities.
\begin{table}[h]
\begin{center}
\begin{tabular}{|c||c|c||c|c|}
\hline
$\sqrt{s}=10$ TeV & \multicolumn{2}{|c||}{$pp\rightarrow e^+ e^-$} & \multicolumn{2}{|c|}{$pp\rightarrow \mu^+ \mu^-$} \\
\hline
$\int \mathcal{L}$ (fb$^{-1}$)  & $g'_1=0.1$ & $g'_1=0.2$  & $g'_1=0.1$ & $g'_1=0.2$ \\
\hline
0.3 & 1100(900)  & 1600(-)    & 1000(800) & 1500(-)       \\  
0.5 & 1250(1000) & 1750(1400) & 1150(900) & 1650(-)      \\ 
1   & 1450(1200) & 1950(1600) & 1350(1100)& 1850(1550)    \\ 
\hline
\end{tabular}
\end{center}
\vskip -0.5cm
\caption{\it Maximum $Z'_{B-L}$ boson masses (in GeV) for a $3\sigma$($5\sigma$) discovery for selected $g_1'$ and integrated luminosities in the $B-L$ model. No numbers are quoted for already excluded configurations.}
\label{5sigma_at_10TeV}
\end{table}

\begin{figure}[!h]
  \subfloat[]{ 
  \label{10TeV_3sigma_el}
  \includegraphics[angle=0,width=0.48\textwidth ]{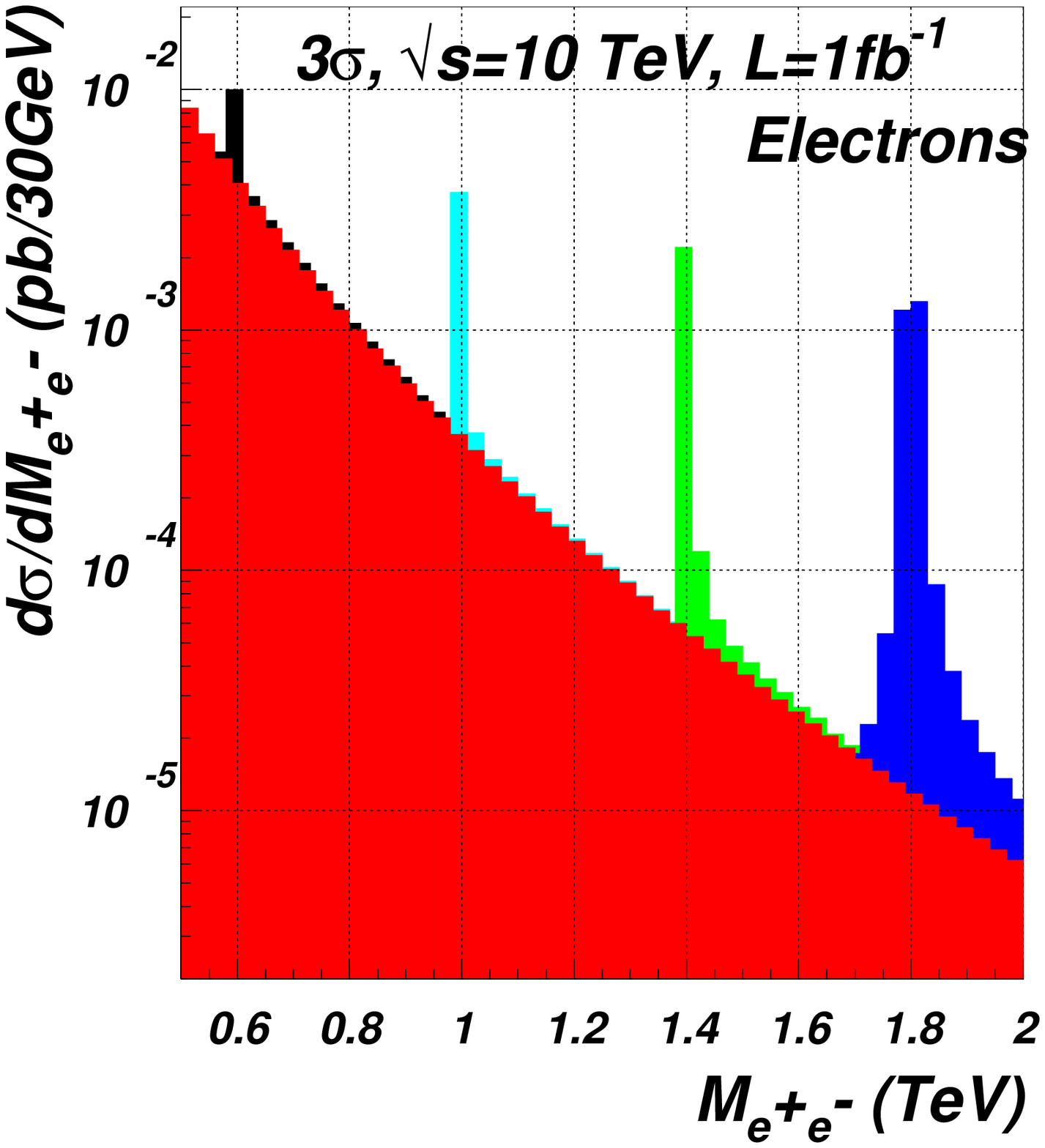}}
  \subfloat[]{
  \label{10TeV_5sigma_el}
  \includegraphics[angle=0,width=0.48\textwidth ]{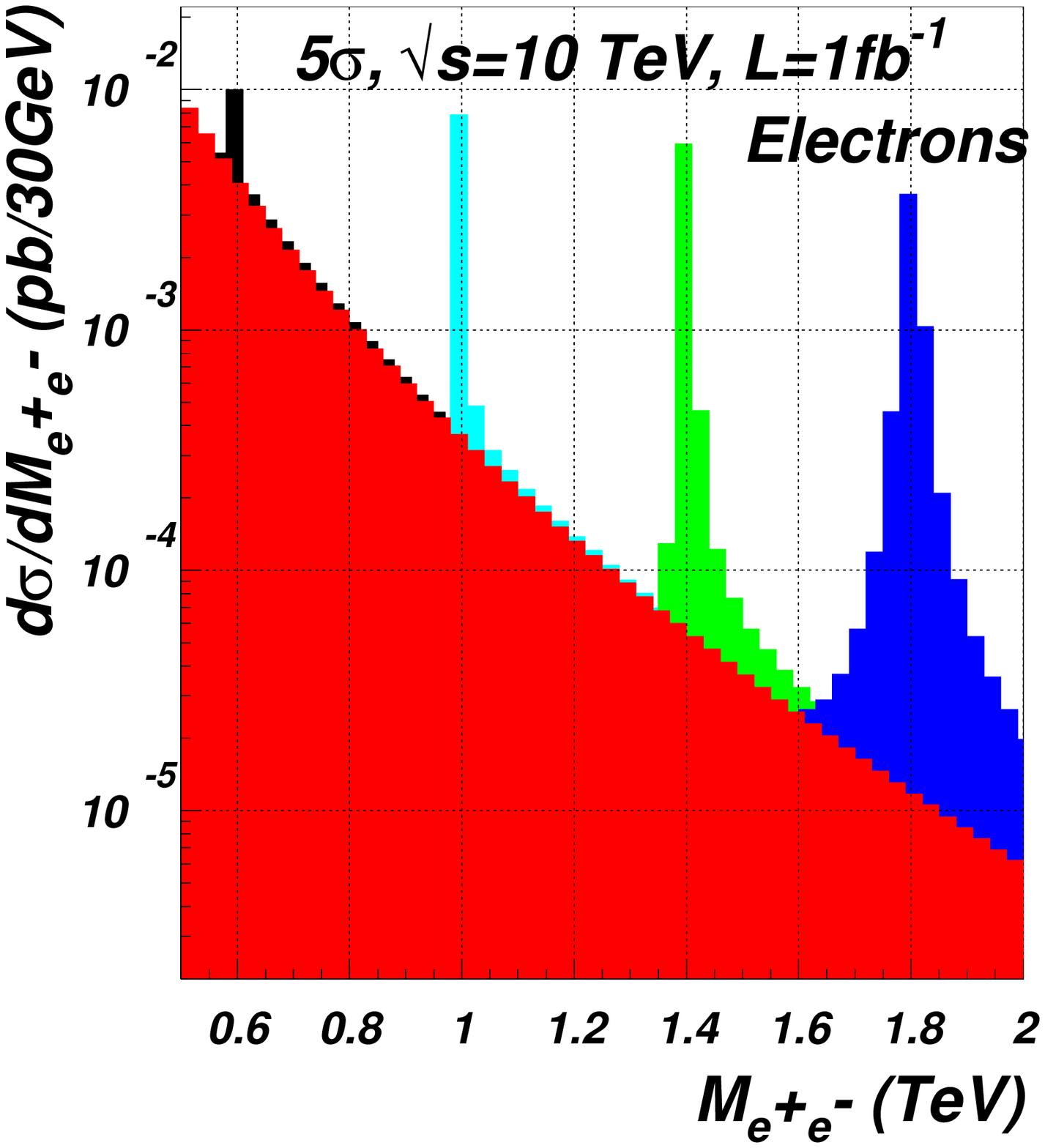}}\\
  \subfloat[]{ 
  \label{10TeV_3sigma_mu}
  \includegraphics[angle=0,width=0.48\textwidth ]{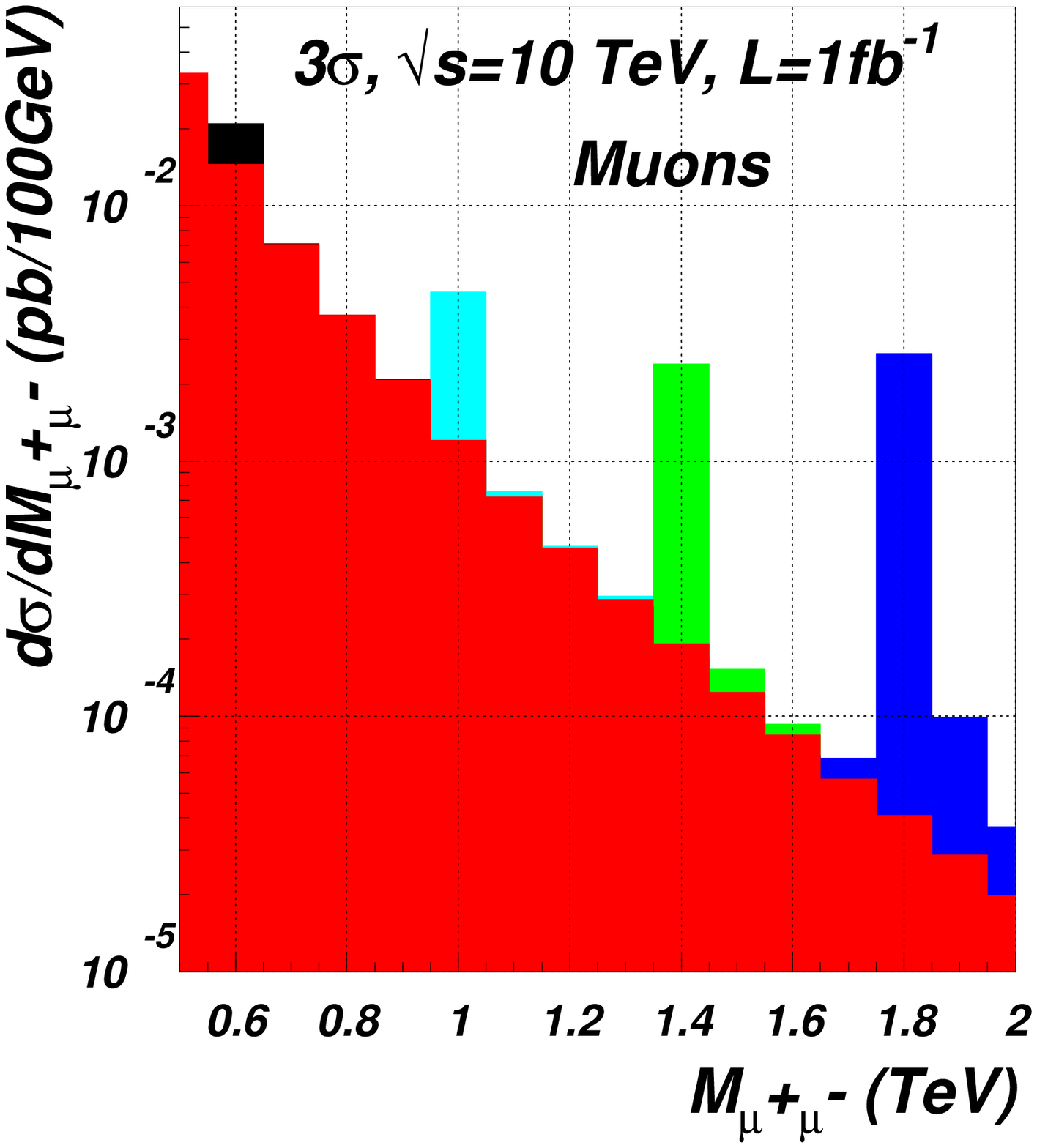}}
  \subfloat[]{
  \label{10TeV_5sigma_mu}
  \includegraphics[angle=0,width=0.48\textwidth ]{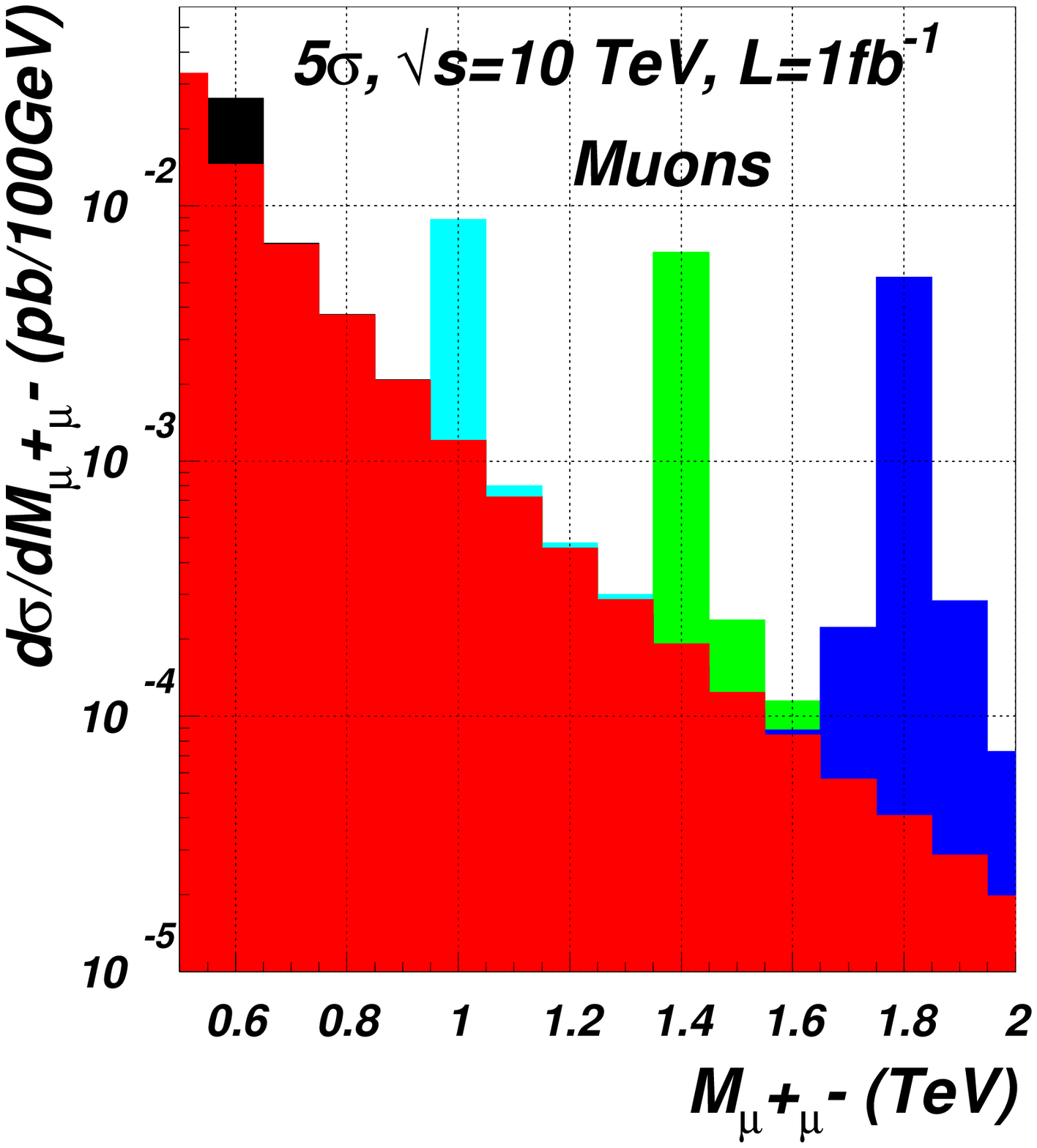}}
  \vspace*{-0.5cm}
  \caption{\it $\frac{d\sigma}{dM_{ll}}(pp\rightarrow \gamma ,Z,Z'_{B-L} \rightarrow \ell ^+\ell ^-)$ for several masses and couplings ($M_{Z'}/TeV$, $g'_1$, $\Gamma _{Z'}/GeV$): ($0.6$, $0.02$, $0.045$), ($1.0$, $0.04$, $0.51$), ($1.4$, $0.07$, $1.4$) and ($1.8$, $0.15$, $8.5$) for  (\ref{10TeV_3sigma_el}) $\ell =e$ and (\ref{10TeV_3sigma_mu}) $\ell =\mu$, corresponding at roughly $3\sigma$; ($0.6$, $0.027$, $0.083$), ($1.0$, $0.06$, $0.73$), ($1.4$, $0.12$, $4.2$) and ($1.8$, $0.22$, $18$) for (\ref{10TeV_5sigma_el}) $\ell =e$ and (\ref{10TeV_5sigma_mu}) $\ell =\mu$, corresponding at roughly $5\sigma$, both from the corresponding lines at $1$ fb$^{-1}$ of figure~\ref{contour10}. Notice that the asymmetry of the peaks is the result of our choice to consider the full interference structure. ($\sqrt{s}=10$ TeV; 30 GeV and 100 GeV binning for electrons and muons, respectively.)}
  \label{10TeV_3-5sigma}
\end{figure}

For illustrative purposes, we choose several benchmark points on the $3\sigma$ and $5\sigma$ lines for $1$ fb$^{-1}$ of data at $\sqrt{s}=10$ TeV of figure~\ref{contour10} and plot the dielectron and dimuon invariant masses, applying the cuts of eq.~(\ref{LHC_cut}) (without selecting any mass window). Figure~\ref{10TeV_3-5sigma} shows the results in this case, for electrons and muons in the final state, respectively, with typical energy resolutions chosen as binning in the plots. Although in any case the peak widths will be dominated by the experimental resolution (quite poor in the early stages of the LHC, especially for muons), only in the case of electrons and for high masses one could appreciate the Breit-Wigner shape of the peak, which instead will appear as a single bin for most of the parameter space along the $3\sigma$ and $5\sigma$ lines if looking at muons in the final state. Bare in mind though that this statement holds just for the minimal conditions, for sets of masses and couplings that minimally allow for the observation/discovery of the $Z'_{B-L}$ gauge boson. As the width increases quadratically with the coupling $g'_1$ (for fixed masses), it is expected that for significances bigger than the lowest appreciable one the peak will broaden significantly, especially for large masses. It is also expected that the width could be accessed for smaller masses by using electrons in the final state, as their resolution is much better than that of muons, thus allowing the natural width to show up at smaller values of the couplings (at fixed mass).


\subsection{LHC at $\boldsymbol{\sqrt{s}=14}$ TeV}
\begin{figure}[!h]
  \subfloat[]{ 
  \label{contour14_ecal}
  \includegraphics[angle=0,width=0.48\textwidth ]{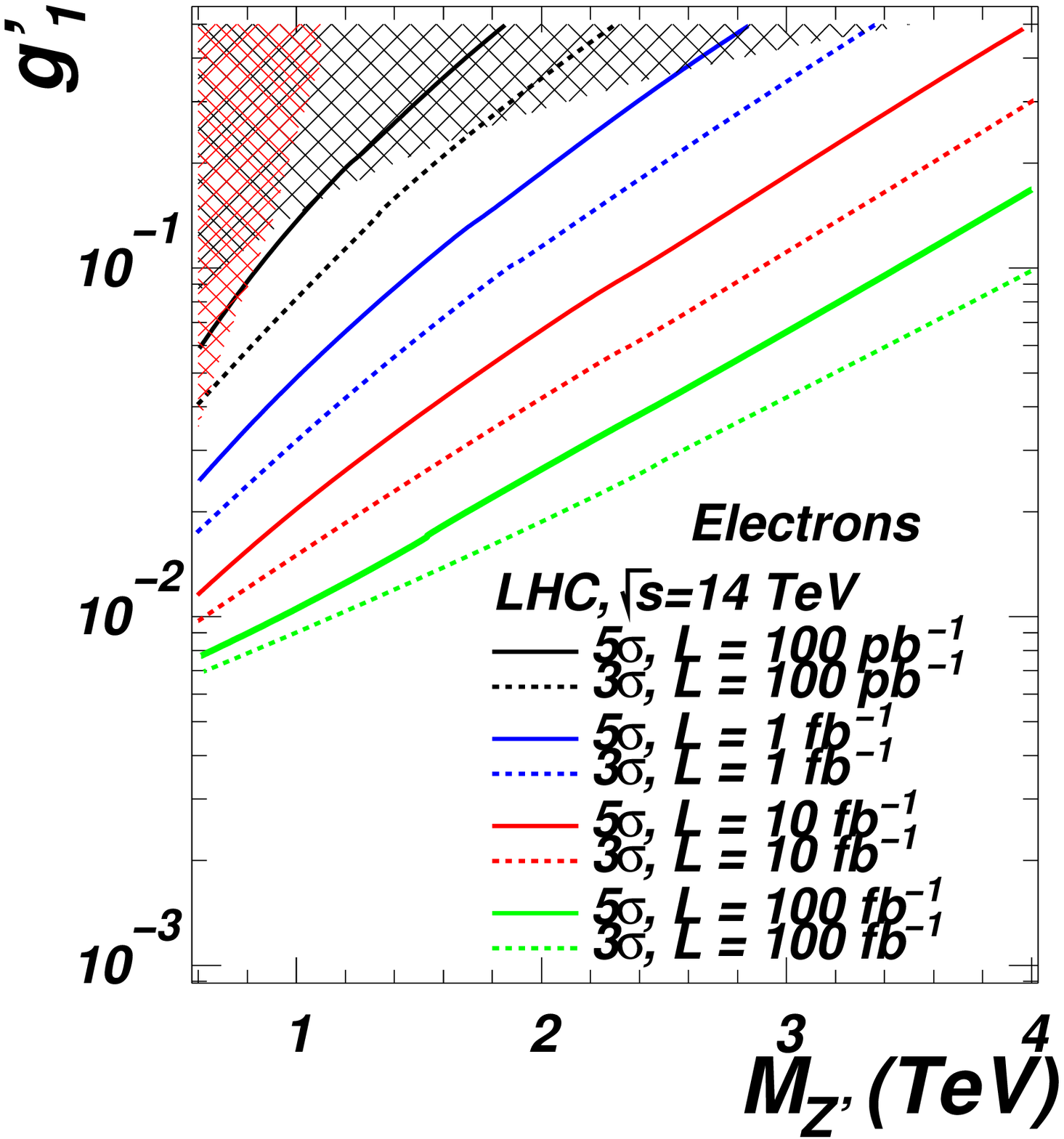}}
  \subfloat[]{
  \label{contour14_pt}
  \includegraphics[angle=0,width=0.48\textwidth ]{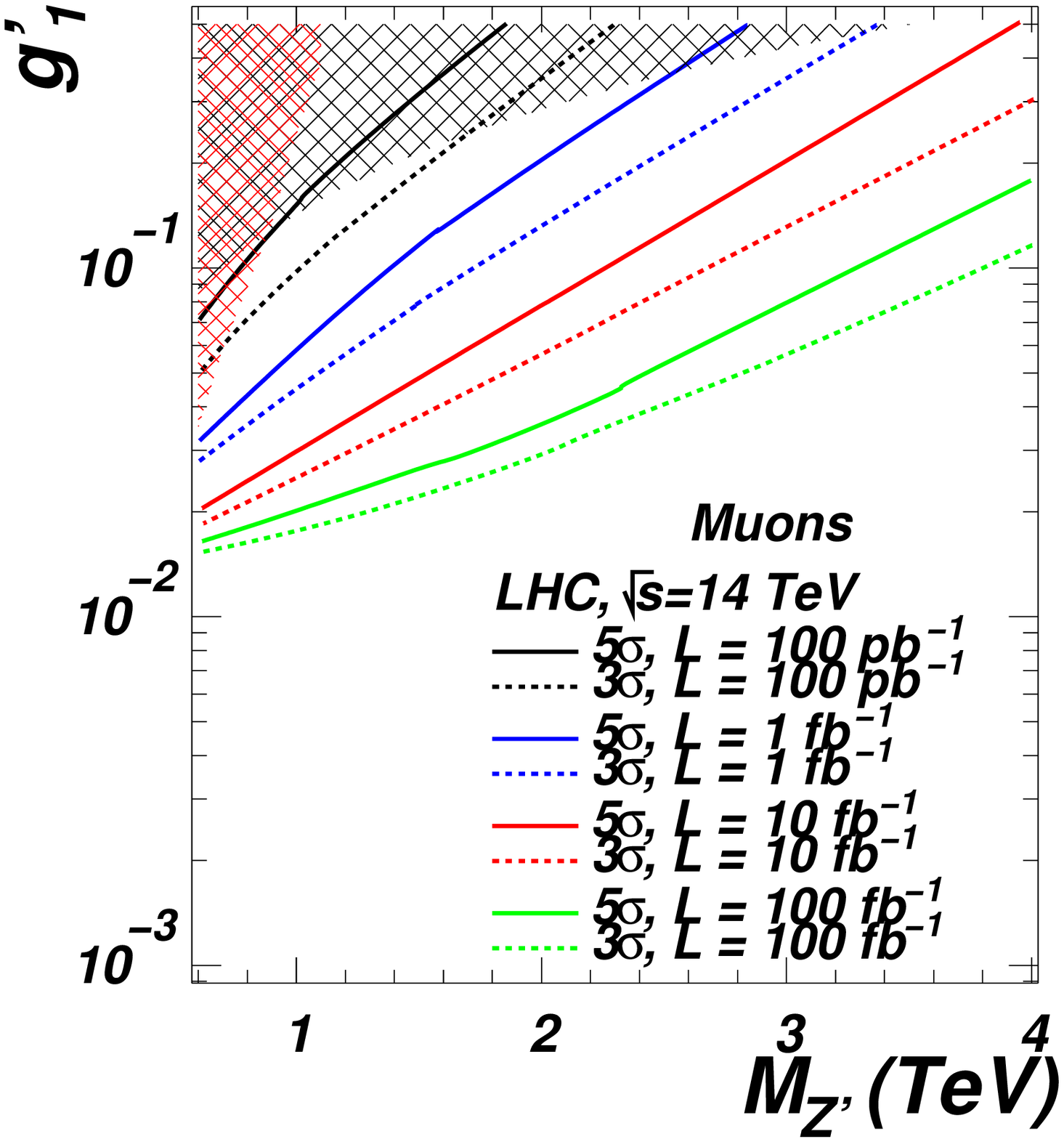}}
  \vspace*{-0.5cm}
  \caption{\it Significance contour levels plotted against $g_1'$
and $M_{Z'}$ at the LHC for $\sqrt{s}=14$ TeV for several integrated luminosities for (\ref{contour14_ecal}) electrons and (\ref{contour14_pt}) muons. The shaded areas correspond to the region of parameter space excluded
experimentally, in accordance with eq.~(\ref{LEP_bound}) (LEP bounds, in black) and table~\ref{mzp-low_bound} (Tevatron bounds, in red).}
  \label{contour14}
\end{figure}

\begin{figure}[!h]
  \subfloat[]{ 
  \label{contour14_ecal_Lumi}
  \includegraphics[angle=0,width=0.48\textwidth ]{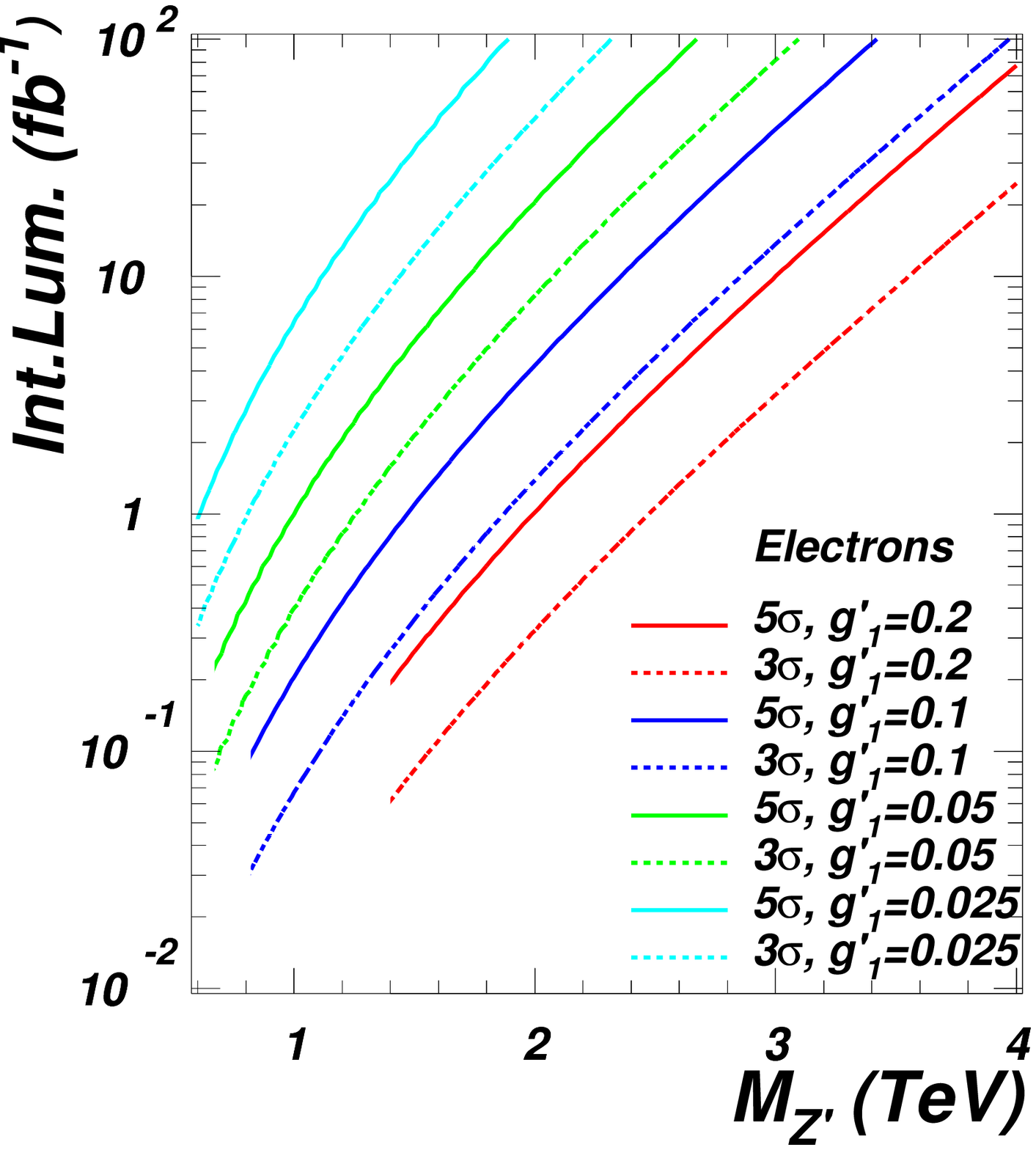}}
  \subfloat[]{
  \label{contour14_pt_Lumi}
  \includegraphics[angle=0,width=0.48\textwidth ]{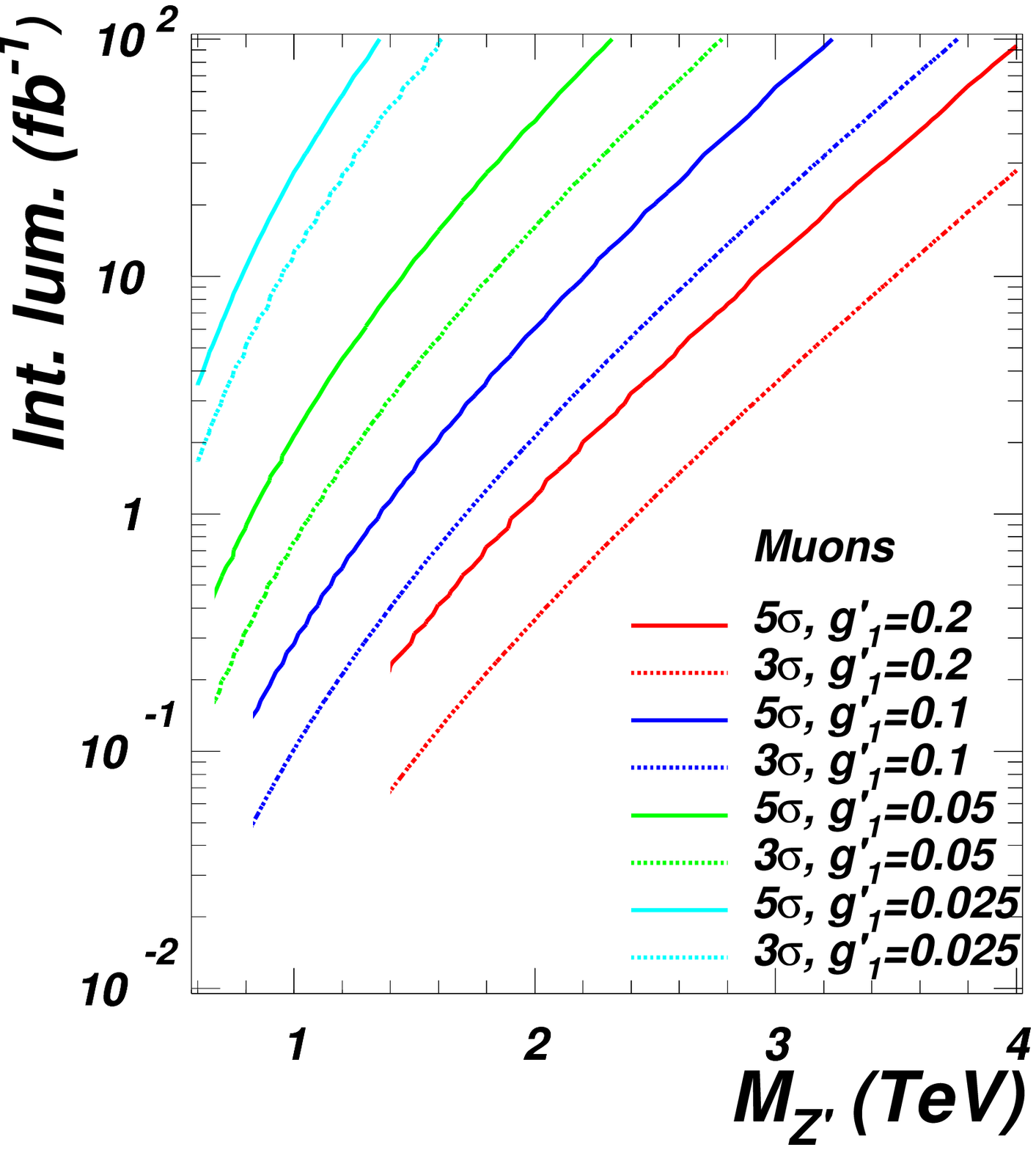}}
  \vspace*{-0.5cm}
  \caption{\it Integrated luminosity required for observation at $3\sigma$ and $5\sigma$ vs. $M_{Z'}$ for selected values of $g_1'$ at the LHC for $\sqrt{s}=14$ TeV for (\ref{contour14_ecal_Lumi}) electrons and (\ref{contour14_pt_Lumi}) muons. Only allowed combination of masses and couplings are shown, in accordance with eq.~(\ref{LEP_bound}) and table~\ref{mzp-low_bound}.}
  \label{lumi_vs_mzp_14TeV}
\end{figure}

We consider here the design performance, i.e., $\sqrt{s}=14$ TeV of CM energy with large luminosity, $\int \mathcal{L} = 100$ fb$^{-1}$. Figure~\ref{contour14} show the discovery potential for the $Z'_{B-L}$ boson under these conditions, while figure~\ref{lumi_vs_mzp_14TeV} show the integrated luminosity required for $3\sigma$ evidence as well as for $5\sigma$ discovery as a function of the $Z'_{B-L}$ boson mass for selected values of the coupling at $\sqrt{s}=14$ TeV. We consider the integrated luminosity in the range between $10$ pb$^{-1}$ up to $100$ fb$^{-1}$. After some years of data analysis, the performances of the detector will be better understood. We therefore use the resolutions for both electrons and muons quoted in eqs.~(\ref{LHC_ris_el_imp}) and (\ref{LHC_ris_mu_imp}), respectively.

From figures~\ref{contour10} and \ref{contour14}, we can see that the LHC at $\sqrt{s}=14$ TeV will start probing a completely new region of the parameter space for $\int \mathcal{L} \geq 1$ fb$^{-1}$.
For $\int \mathcal{L} \geq 10$ fb$^{-1}$ a $Z'_{B-L}$ gauge boson can be discovered up to masses of $4$ TeV and for couplings as small as $0.01(0.02)$ if we are dealing with electrons(muons). At $\int \mathcal{L} = 100$ fb$^{-1}$, the coupling can be probed down to values of $8\,\cdot 10^{-3}$ in the electron channel, while couplings smaller than $1.6\,\cdot 10^{-2}$ cannot be accessed with muons. The mass region that can be covered extends towards $5$ TeV irrespectively of the final state.

As before, figure~\ref{lumi_vs_mzp_14TeV} shows the integrated luminosity required for $3(5)\sigma$ evidence(discovery) of the $Z'_{B-L}$ boson as a function of
the mass, for selected values of the coupling. We explore the range in luminosities, from $10$ pb$^{-1}$ to $100$ fb$^{-1}$. However, just the configuration with $g'_1=0.1$ can be probed with very low luminosity, requiring $30(100)$ pb$^{-1}$ and $50(150)$ pb$^{-1}$ at $3\sigma(5\sigma)$ respectively considering electrons and muons in the final state. For values of the coupling such as $0.05$ and $0.2$, $90(220)$ pb$^{-1}$ and $60(200)$ pb$^{-1}$ are the integrated luminosities required to start to be sensitive (at $3(5)\sigma$) if electrons are considered, while $160(500)$ and $70(220)$ pb$^{-1}$ are the least integrated luminosity required, respectively, if instead we look at muons. It is worth to emphasise here that the first couplings that will start to be probed at the LHC are those around $g'_1=0.1$.

The better resolution in the case of electrons reflects in a better sensitivity to smaller $Z'_{B-L}$ masses with respect to muons. For $M_{Z'}=600$ GeV, the LHC with $\sqrt{s}=14$ TeV requires $1.0$ fb$^{-1}$ to be sensitive at $5\sigma$ to a value of the coupling of $0.025$ in the electron channel. If we are considering muons, $3.5$ fb$^{-1}$ is the required luminosity to probe at $5\sigma$ the same value of the coupling.

The $5\sigma$ discovery potential for the LHC at $\sqrt{s}=14$ TeV is summarised in table~\ref{5sigma_at_14TeV}, for selected values of $Z'_{B-L}$ masses and couplings.
\begin{table}[h]
\begin{center}
\begin{tabular}{|c||c|c|c||c|c|c|}
\hline
$\sqrt{s}=14$ TeV & \multicolumn{3}{|c||}{$pp\rightarrow e^+ e^-$} & \multicolumn{3}{|c|}{$pp\rightarrow \mu^+ \mu^-$} \\
\hline
$g'_1$ & $M_{Z'}=1$ TeV & $M_{Z'}=2$ TeV  & $M_{Z'}=3$ TeV & $M_{Z'}=1$ TeV & $M_{Z'}=2$ TeV  & $M_{Z'}=3$ TeV \\
\hline
0.025 & 2.5(7.0)  & 50($>$100)& $>$100($>$300)& 15(30)  & $>$100($>$100)& $>$300($>$300) \\
0.05  & 0.4(1.0)  &  9(20)    & 80($>$100)    & 0.8(2.5)&   20(50)      & $>$100($>$100)\\
0.1   & 0.07(0.2) & 1.5(4.0)  & 15(50)        & 0.1(0.3)&   2.0(6.0)    &  20(65)  \\
0.2   & $-$($-$)  & 0.3(1.0)  & 3(10)         & $-$($-$)&   0.4(1.2)    &  3(12)\\
\hline  
\end{tabular}
\end{center}
\vskip -0.5cm
\caption{\it Minimum integrated luminosities (in fb$^{-1}$) for a $3\sigma$($5\sigma$) discovery for selected $Z'_{B-L}$ boson masses and $g_1'$ couplings for the $B-L$ model. No numbers are quoted for already excluded configurations.}
\label{5sigma_at_14TeV}
\end{table}

\begin{figure}[!h]
  \subfloat[]{ 
  \label{14TeV_3sigma_el}
  \includegraphics[angle=0,width=0.48\textwidth ]{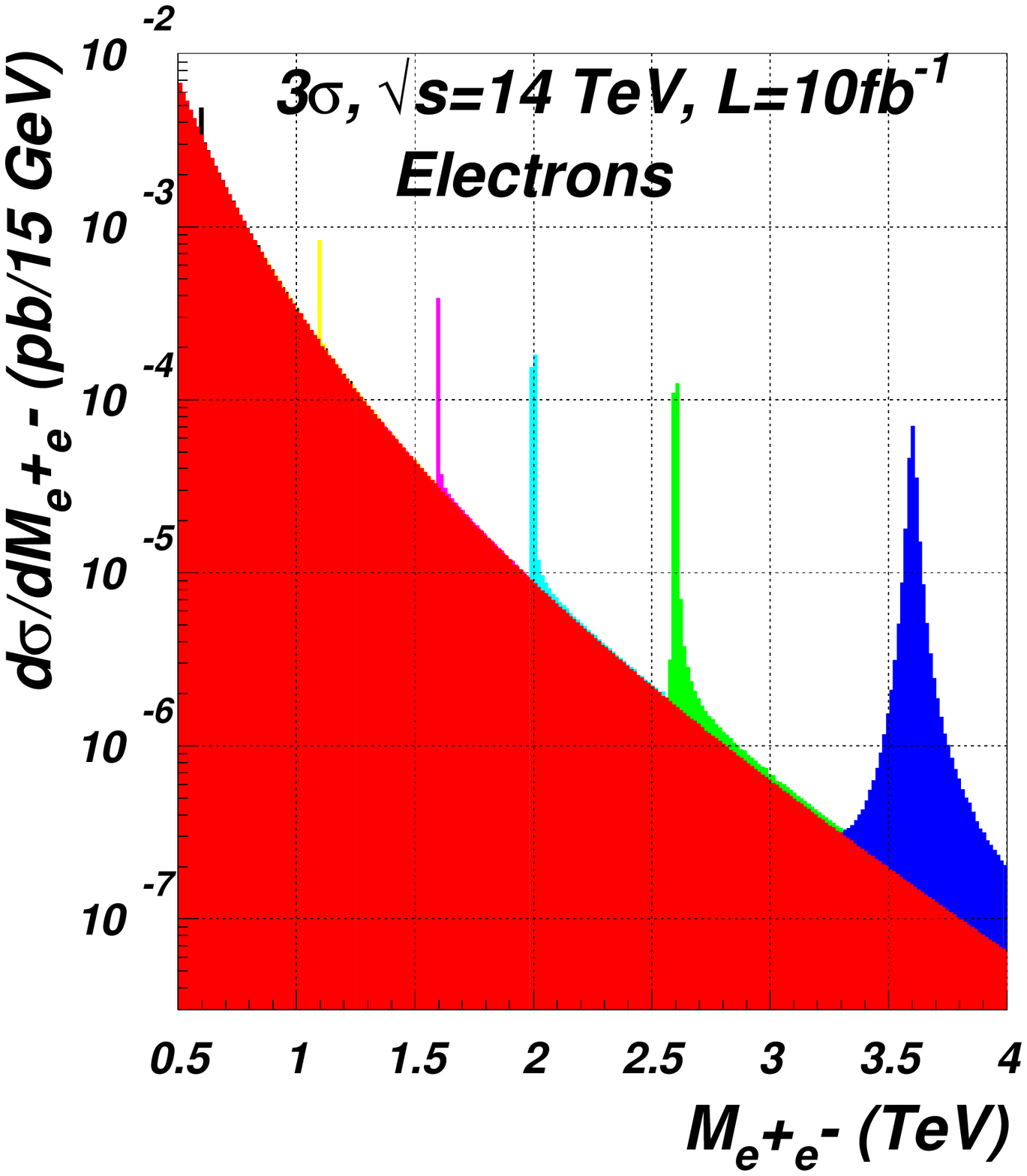}}
  \subfloat[]{
  \label{14TeV_5sigma_el}
  \includegraphics[angle=0,width=0.48\textwidth ]{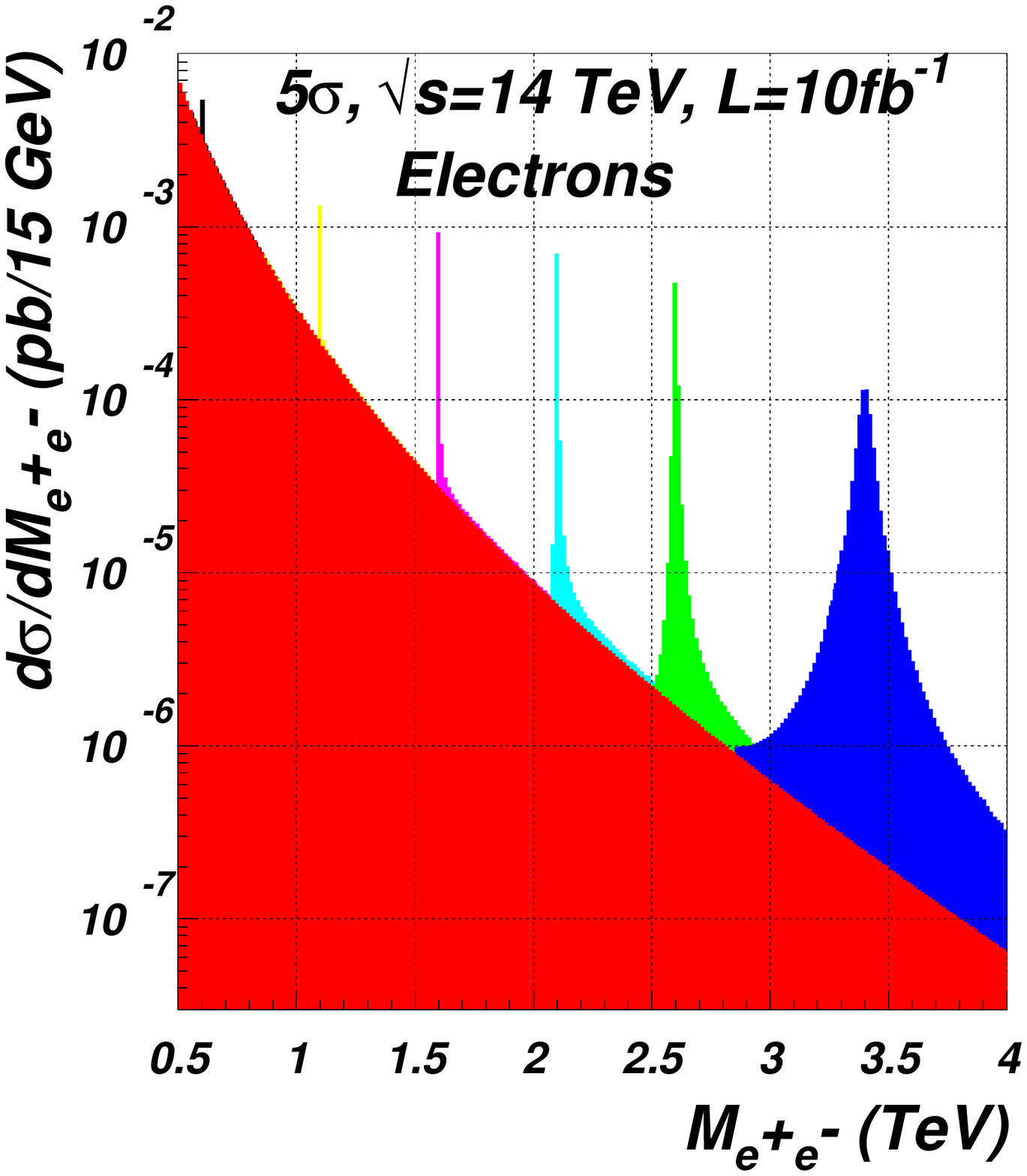}}
  \vspace*{-0.5cm}
  \caption{\it $\frac{d\sigma}{dM_{ll}}(pp\rightarrow \gamma ,Z,Z'_{B-L} \rightarrow e ^+e ^-)$ for several masses and couplings ($M_{Z'}/$TeV, $g'_1$, $\Gamma _{Z'}/GeV$): (\ref{14TeV_3sigma_el}) ($0.6$, $0.0075$, $0.006$), ($1.1$, $0.015$, $0.05$), ($1.6$, $0.025$, $0.21$), ($2.0$, $0.04$, $0.67$), ($2.6$, $0.07$, $2.7$) and ($3.6$, $0.2$, $31$); (\ref{14TeV_5sigma_el}) ($0.6$, $0.009$, $0.009$), ($1.1$, $0.02$, $0.09$), ($1.6$, $0.04$, $0.53$), ($2.1$, $0.07$, $2.2$), ($2.6$, $0.12$, $7.9$) and ($3.4$, $0.3$, $61$), from the $3\sigma$ and $5\sigma$ lines at $10$ fb$^{-1}$ of figure~\ref{contour14} ($\sqrt{s}=14$ TeV), respectively, using 15 GeV binning. Notice that the asymmetry of the peaks is the result of our choice to consider the full interference structure.}
  \label{14TeV_3-5sigma_el}
\end{figure}

\begin{figure}[!h]
  \subfloat[]{ 
  \label{14TeV_3sigma_mu}
  \includegraphics[angle=0,width=0.48\textwidth ]{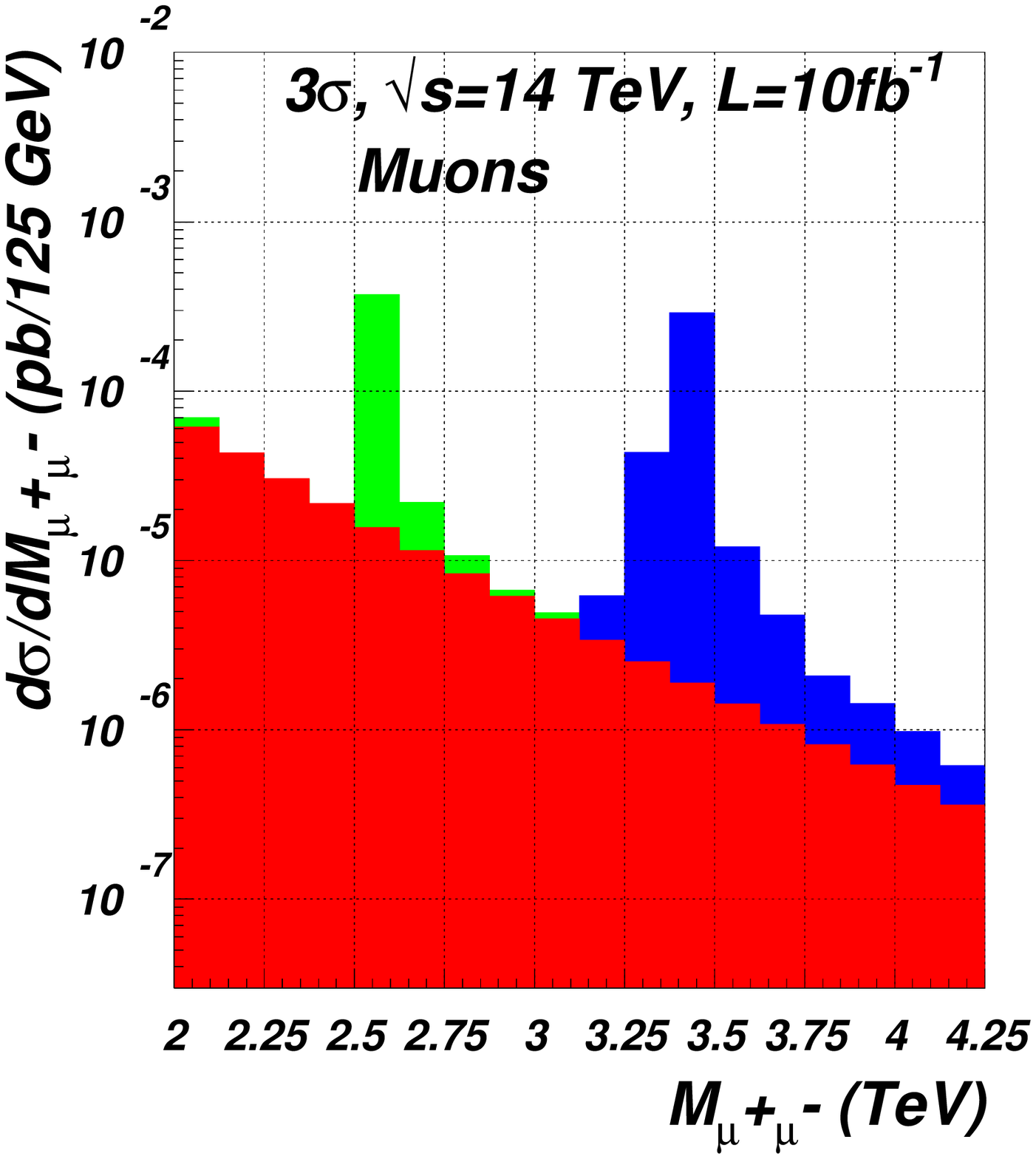}}
  \subfloat[]{
  \label{14TeV_5sigma_mu}
  \includegraphics[angle=0,width=0.48\textwidth ]{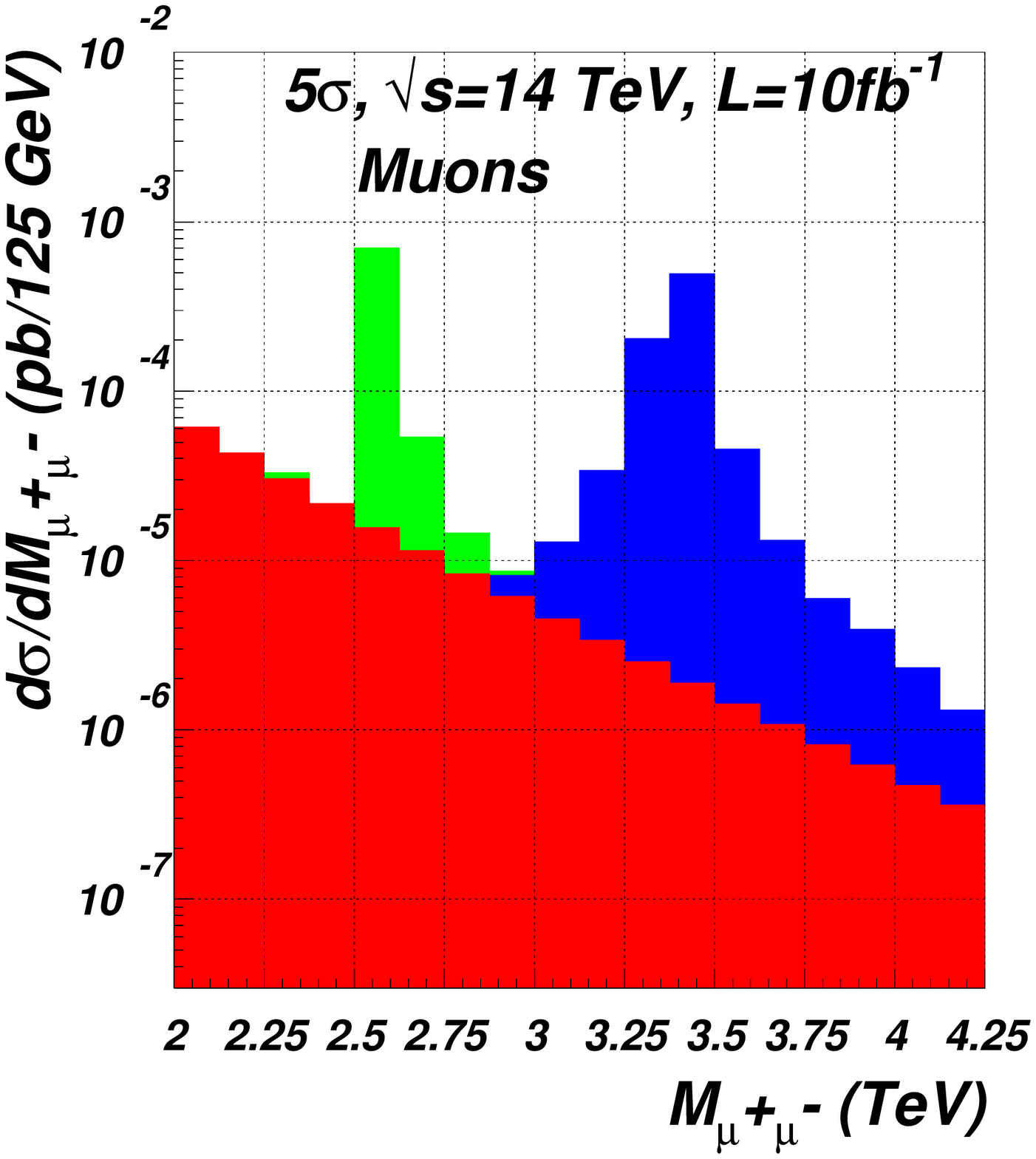}}
  \vspace*{-0.5cm}
  \caption{\it $\frac{d\sigma}{dM_{ll}}(pp\rightarrow \gamma ,Z,Z' \rightarrow \mu ^+\mu ^-)$ for some masses and couplings ($M_{Z'}/$TeV, $g'_1$, $\Gamma _{Z'}/GeV$): (\ref{14TeV_3sigma_mu}) ($2.6$, $0.07$, $2.7$) and ($3.4$, $0.2$, $29$) and (\ref{14TeV_5sigma_mu}) ($2.6$, $0.12$, $8$) and ($3.4$, $0.3$, $65$), from the $3\sigma$ and $5\sigma$ lines at $10$ fb$^{-1}$ of figure~\ref{contour14} ($\sqrt{s}=14$ TeV), using $125$ GeV binning. Notice that the asymmetry of the peaks is the result of our choice to consider the full interference structure. }
  \label{14TeV_3-5sigma_mu}
\end{figure}

Again, figures~\ref{14TeV_3-5sigma_el} and \ref{14TeV_3-5sigma_mu} show a pictorial representation of the $Z'$ properties (widths and cross sections) for selected benchmark points on the $3\sigma$ and $5\sigma$ lines for $10$ fb$^{-1}$ of data at $\sqrt{s}=14$ TeV, plotting the dilepton invariant mass to which just the cuts of eq.~(\ref{LHC_cut}) have been applied (without selecting any mass window).

The improved resolution for electrons allows a measure of the $Z'_{B-L}$ boson width not only at high masses, but also opens the possibility of a measurement even for smaller masses as compared to the $\sqrt{s}=10$ TeV setup.

\section{$Z'$ sector: exclusion power}\label{sect:Zp_excl}
If no evidence for a signal is found at the LHC at any energy and luminosity configurations, $95\%$ C.L. exclusion limits can be derived: in the following subsections we present exclusion plots for each stage of the LHC CM energy. We also show the expected exclusions at the Tevatron for $\int{\mathcal{L}}=10$ fb$^{-1}$.

\subsection{LHC at $\boldsymbol{\sqrt{s}=7}$ TeV}

\begin{figure}[!h]
  \subfloat[]{ 
  \label{contour7_excl}
  \includegraphics[angle=0,width=0.48\textwidth ]{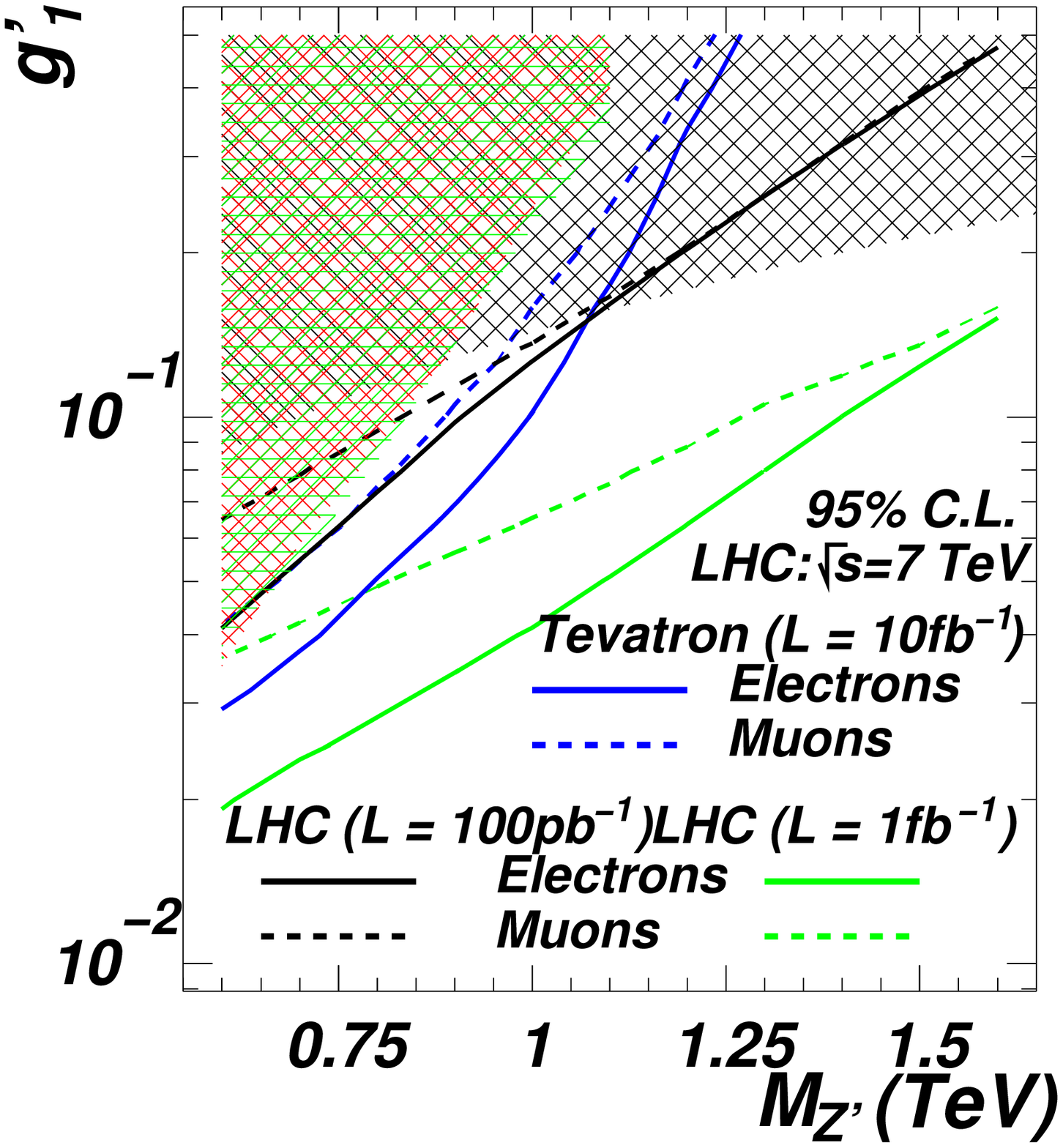}}\\
  \subfloat[]{
  \label{lumi7LHC_excl}
  \includegraphics[angle=0,width=0.48\textwidth ]{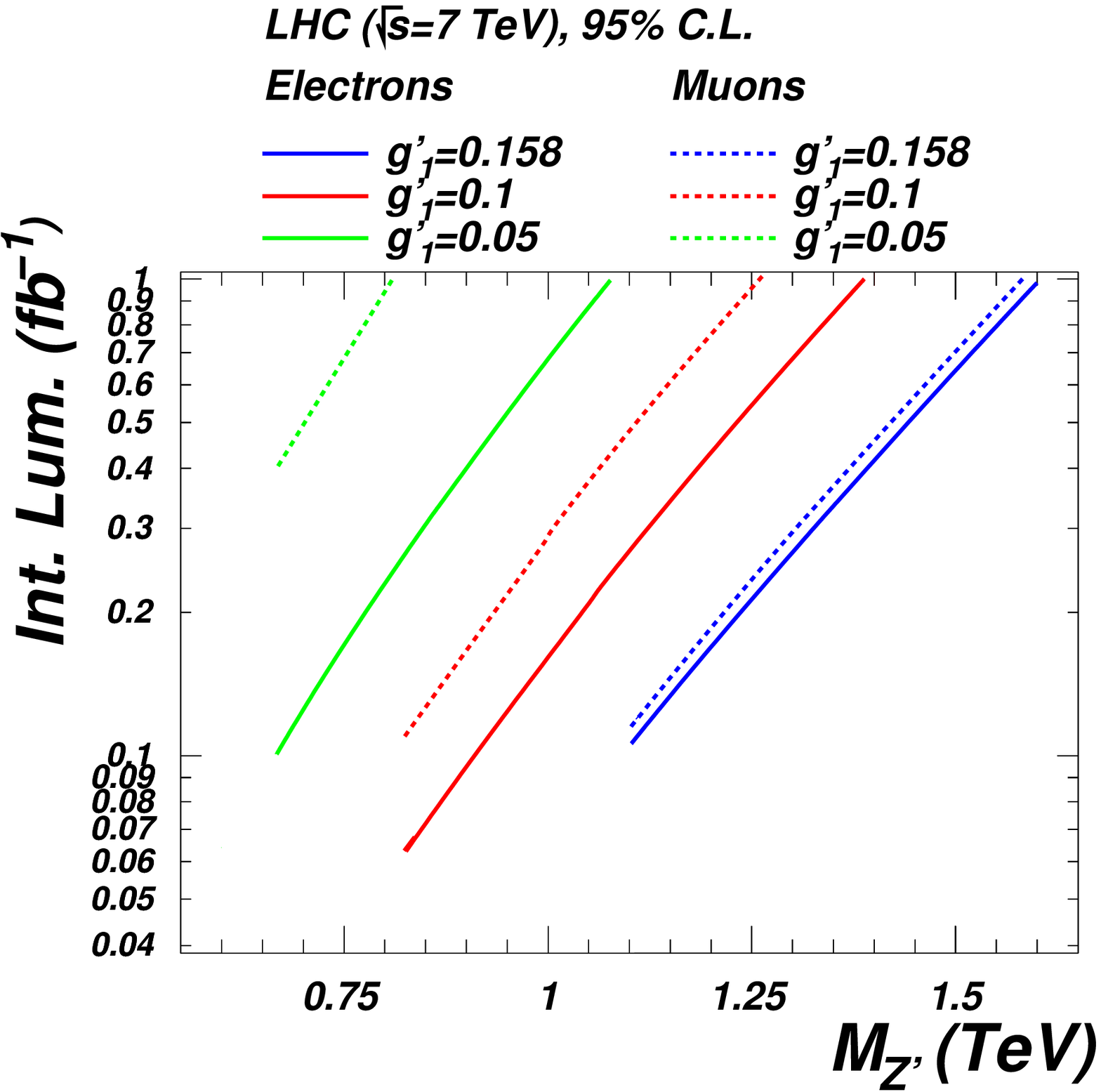}}
  \subfloat[]{
  \label{lumi7Tev_excl}
  \includegraphics[angle=0,width=0.48\textwidth ]{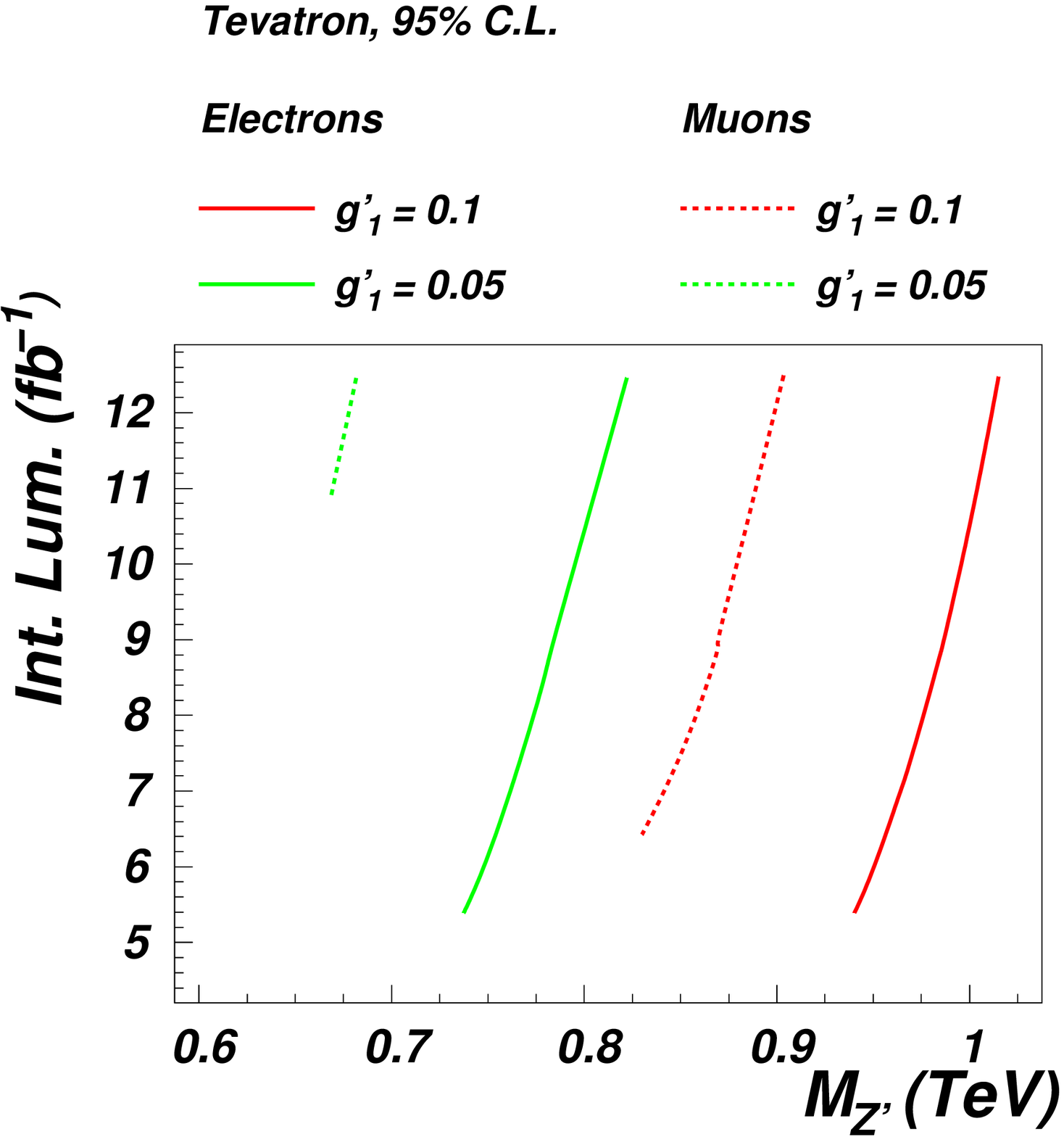}}
  \vspace*{-0.5cm}
  \caption{\it (\ref{contour7_excl}) Contour levels for $95\%$ C.L. plotted against $g_1'$ and $M_{Z'}$ at the LHC for selected integrated luminosities and integrated luminosity required for observation at $3\sigma$ and $5\sigma$ vs. $M_{Z'}$ for selected values of $g_1'$ (in which only the allowed combination of masses and couplings are shown), for (\ref{lumi7LHC_excl}) the LHC at $\sqrt{s}=7$ TeV and (\ref{lumi7Tev_excl}) the Tevatron ($\sqrt{s}=1.96$ TeV), for both electrons and muons.
The shaded areas and the allowed $(M_{Z'},g'_1)$ shown are in accordance with eq.~(\ref{LEP_bound}) (LEP bounds, in black) and table~\ref{mzp-low_bound} (Tevatron bounds, in red for electrons and in green for muons).}
  \label{excl_7}
\end{figure}

We start by looking at the $95\%$ C.L. limits presented in figure~\ref{excl_7} for the Tevatron and for this stage of the LHC (for $10$ fb$^{-1}$ and $1$ fb$^{-1}$ of integrated luminosity, respectively).

One can see that the different resolutions imply that the limits derived using electrons are always more stringent than those derived using muons in excluding the $Z'_{B-L}$ boson.
As for the discovery reach, the Tevatron is also competitive in setting limits, especially in the lower mass region.
In particular, using electrons and in case of no evidence at the Tevatron with $10$ fb$^{-1}$, the $Z'_{B-L}$ boson can be excluded for values of the coupling down to $0.03$ ($0.04$ for muons) for $M_{Z'}=600$ GeV. For the LHC to set the same exclusion limit for the same mass, $1$ fb$^{-1}$ of integrated luminosity is required, allowing to exclude $g'_1>0.02(0.35)$ using electron(muons) in the final state. For the same integrated luminosity, the LHC has much more scope in excluding a $Z'_{B-L}$, for $M_{Z'}>1.0$ TeV.

For a coupling of $0.1$, the $Z'_{B-L}$ boson can be excluded up to $1.40(1.25)$ TeV at the LHC considering electrons(muons) for $1$ fb$^{-1}$, and up to $1.0(0.9)$ TeV at the Tevatron for $10$ fb$^{-1}$ of data. For $g'_1=0.05$, the LHC when looking at muons(electrons) will require $400(100)$ pb$^{-1}$ to start improving the current available limits, while with $100$ pb$^{-1}$ it can set limits on $g'_1=0.158$, out of the reach of Tevatron. It will ultimately be able to exclude $Z'_{B-L}$ boson up to $M_{Z'}=1.6$ TeV for $1$ fb$^{-1}$ (both with electrons and muons).

The $95\%$ C.L. exclusions for the LHC at $\sqrt{s}=7$ TeV and at the Tevatron are summarised in table~\ref{2sigma_at_7TeV}, for selected values of couplings and integrated luminosities.
\begin{table}[h]
\begin{center}
\begin{tabular}{|c||c|c|c||c|c|c|}
\hline
LHC & \multicolumn{3}{|c||}{$pp\rightarrow e^+ e^-$} & \multicolumn{3}{|c|}{$pp\rightarrow \mu^+ \mu^-$} \\
\hline
$\int \mathcal{L}$ (fb$^{-1}$) & $g'_1=0.05$   & $g'_1=0.1$ & $g'_1=0.158$ & $g'_1=0.05$   & $g'_1=0.1$ & $g'_1=0.158$ \\
\hline 
0.1 & 670  & 900  & 1100 & $-$  & 820  & $-$  \\
0.2 & 770  & 1050 & 1250 & $-$  & 950  & 1225  \\
0.5 & 950  & 1225 & 1450 & 700  & 1100 & 1425  \\
  1 & 1075 & 1375 & 1600 & 800  & 1250 & 1575  \\
\hline
\hline
Tevatron & \multicolumn{3}{|c||}{$pp\rightarrow e^+ e^-$} & \multicolumn{3}{|c|}{$pp\rightarrow \mu^+ \mu^-$} \\
\hline
$\int \mathcal{L}$ (fb$^{-1}$) & $g'_1=0.05$   & $g'_1=0.1$ & $g'_1=0.158$ & $g'_1=0.05$   & $g'_1=0.1$ & $g'_1=0.158$ \\
\hline
 6  & 750 & 950  & $-$ & $-$  & $-$ & $-$  \\
 8  & 775 & 975  & $-$ & $-$  & 860 & $-$  \\
 10 & 800 & 1000 & $-$ & $-$  & 875 & $-$  \\
 12 & 825 & 1020 & $-$ & 680  & 900 & $-$  \\
\hline
\end{tabular}
\end{center}
\vskip -0.5cm
\caption{\it Maximum $Z'_{B-L}$ boson masses (in GeV) for a $95\%$ C.L. exclusion for selected $g_1'$ and integrated luminosities in the $B-L$ model. No numbers are quoted for already excluded configurations.}
\label{2sigma_at_7TeV}
\end{table}

\subsection{LHC at $\boldsymbol{\sqrt{s}=10}$ TeV}

\begin{figure}[!h]
  \subfloat[]{ 
  \label{contour10_excl}
  \includegraphics[angle=0,width=0.48\textwidth ]{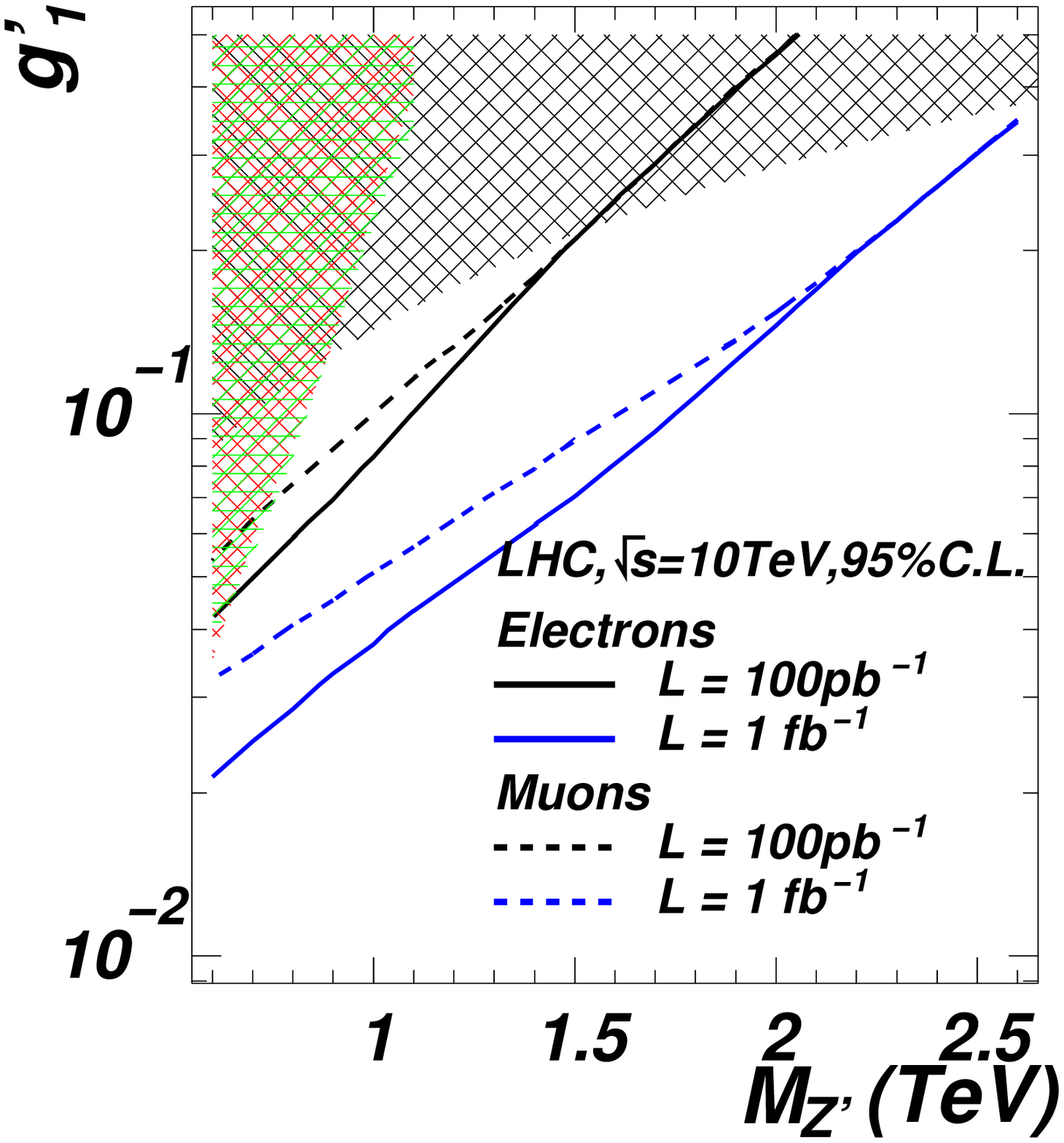}}
  \subfloat[]{
  \label{lumi10_excl}
  \includegraphics[angle=0,width=0.48\textwidth ]{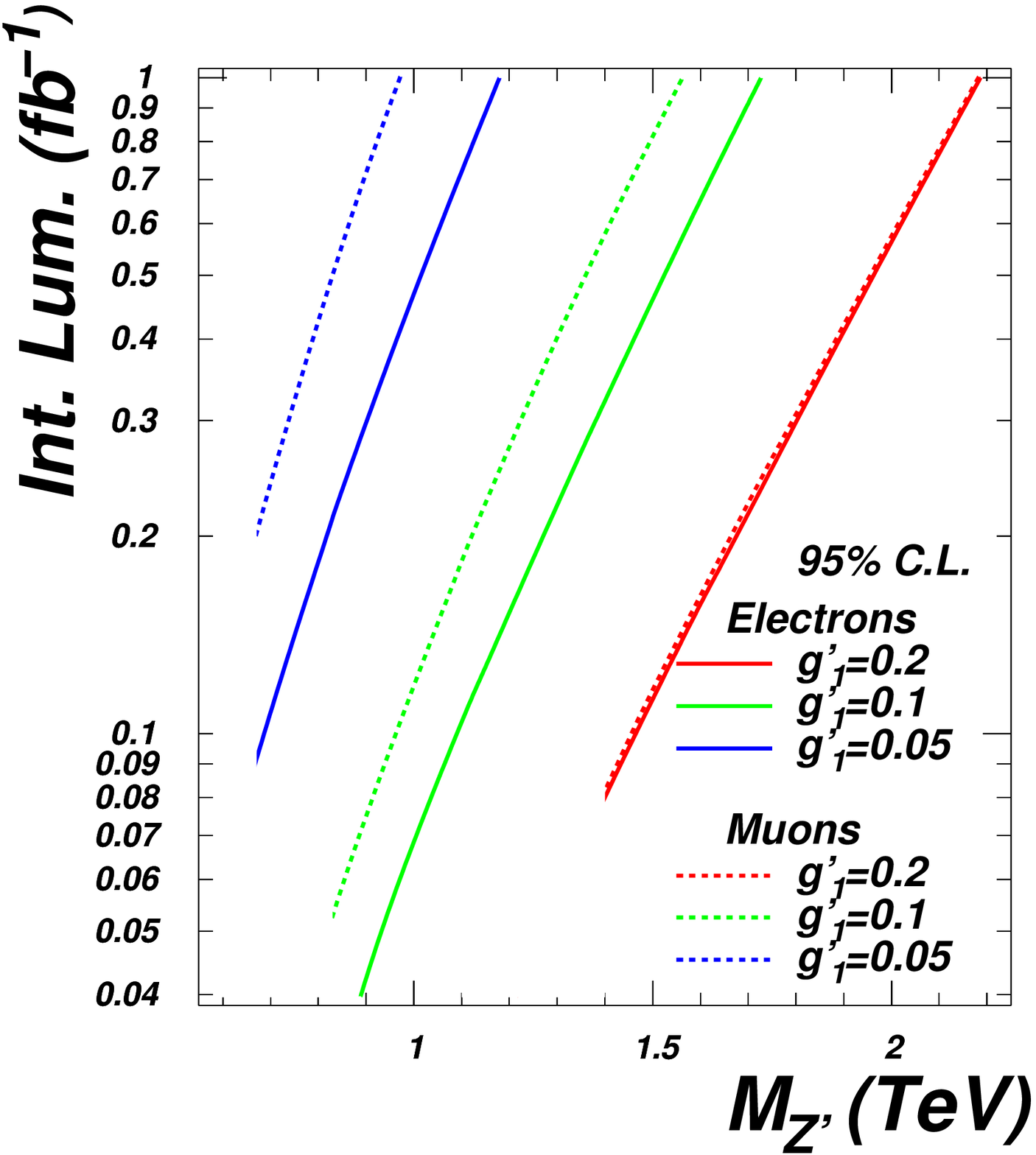}}
  \vspace*{-0.5cm}
  \caption{\it (\ref{contour10_excl}) Contour levels for $95\%$ C.L. plotted against $g_1'$ and $M_{Z'}$ at the LHC for selected integrated luminosities and (\ref{lumi10_excl}) integrated luminosity required for observation at $3\sigma$ and $5\sigma$ vs. $M_{Z'}$ for selected values of $g_1'$ (in which only the allowed combination of masses and couplings are shown), for $\sqrt{s}=10$ TeV, for both electrons
and muons.
The shaded areas and the allowed $(M_{Z'},g'_1)$ shown are in accordance with eq.~(\ref{LEP_bound}) (LEP bounds, in black) and table~\ref{mzp-low_bound} (Tevatron bounds, in red for electrons and in green for muons).}
  \label{excl_10}
\end{figure}

Moving to the LHC at $\sqrt{s}=10$ TeV, supposing that this stage will collect up to $1$ fb$^{-1}$, it would be possible to further extend the region of masses that can be excluded, up to $M_{Z'}=2.5$ TeV irrespectively of whether
one exploits electrons or muons. Figure~\ref{contour10_excl} shows the $95\%$ C.L. exclusions at this stage of the LHC comparing the sensitivities with electrons and muons, for selected values of the integrated luminosity. When $g'_1\gtrsim 0.2$ the two channels set very close limits. 
For smaller couplings, electrons become more sensitive than muons, and for $1$ fb$^{-1}$ both can set more stringent limits than the Tevatron at $10$ fb$^{-1}$, being able to exclude a coupling down to $0.022$ and $0.035$, respectively. Figure~\ref{lumi10_excl} then shows what integrated luminosity is required to exclude a certain $Z'_{B-L}$ boson mass for the CM energy considered here.
As previously noticed, electrons and muons set similar bounds on the mass for a coupling bigger than $\sim 0.2$: here, the far right lines (in red) correspond to $g'_1=0.2$, and this coupling can be excluded from the minimum allowed mass ($1.4$ TeV) up to $2.2$ TeV for $1$ fb$^{-1}$. 
As one decreases the coupling, electrons set more stringent bounds: for $g'_1=0.1(0.05)$, the mass that can be excluded ranges between $900(700)$ GeV and $1.70(1.20)$ TeV for electrons while for muons it ranges between $800(700)$ and $1.55(0.95)$ for the same values of the coupling, the maximum values being for $1$ fb$^{-1}$ of integrated luminosity.

The $95\%$ C.L. exclusions for the LHC at $\sqrt{s}=10$ TeV are summarised in  table~\ref{2sigma_at_10TeV}, for selected values of couplings and integrated luminosities.
\begin{table}[h]
\begin{center}
\begin{tabular}{|c||c|c|c||c|c|c|}
\hline
$\sqrt{s}=10$ TeV & \multicolumn{3}{|c||}{$pp\rightarrow e^+ e^-$} & \multicolumn{3}{|c|}{$pp\rightarrow \mu^+ \mu^-$} \\
\hline
$\int \mathcal{L}$ (fb$^{-1}$) & $g'_1=0.05$   & $g'_1=0.1$ & $g'_1=0.2$ & $g'_1=0.05$   & $g'_1=0.1$ & $g'_1=0.2$ \\
\hline
0.1 & 700  & 1100 & 1450& $-$  & 950  & 1450  \\
0.2 & 800  & 1250 & 1650& 650  & 1100 & 1650  \\
0.5 & 1000 & 1500 & 1950& 800  & 1350 & 1950  \\
  1 & 1150 & 1700 & 2200& 950  & 1550 & 2200  \\
\hline
\end{tabular}
\end{center}
\vskip -0.5cm
\caption{\it Maximum $Z'_{B-L}$ boson masses (in GeV) for a $95\%$ C.L. exclusion for selected $g_1'$ and integrated luminosities in the $B-L$ model. No numbers are quoted for already excluded configurations.}
\label{2sigma_at_10TeV}
\end{table}

\subsection{LHC at $\boldsymbol{\sqrt{s}=14}$ TeV}

\begin{figure}[!h]
  \subfloat[]{ 
  \label{contour14_excl}
  \includegraphics[angle=0,width=0.48\textwidth ]{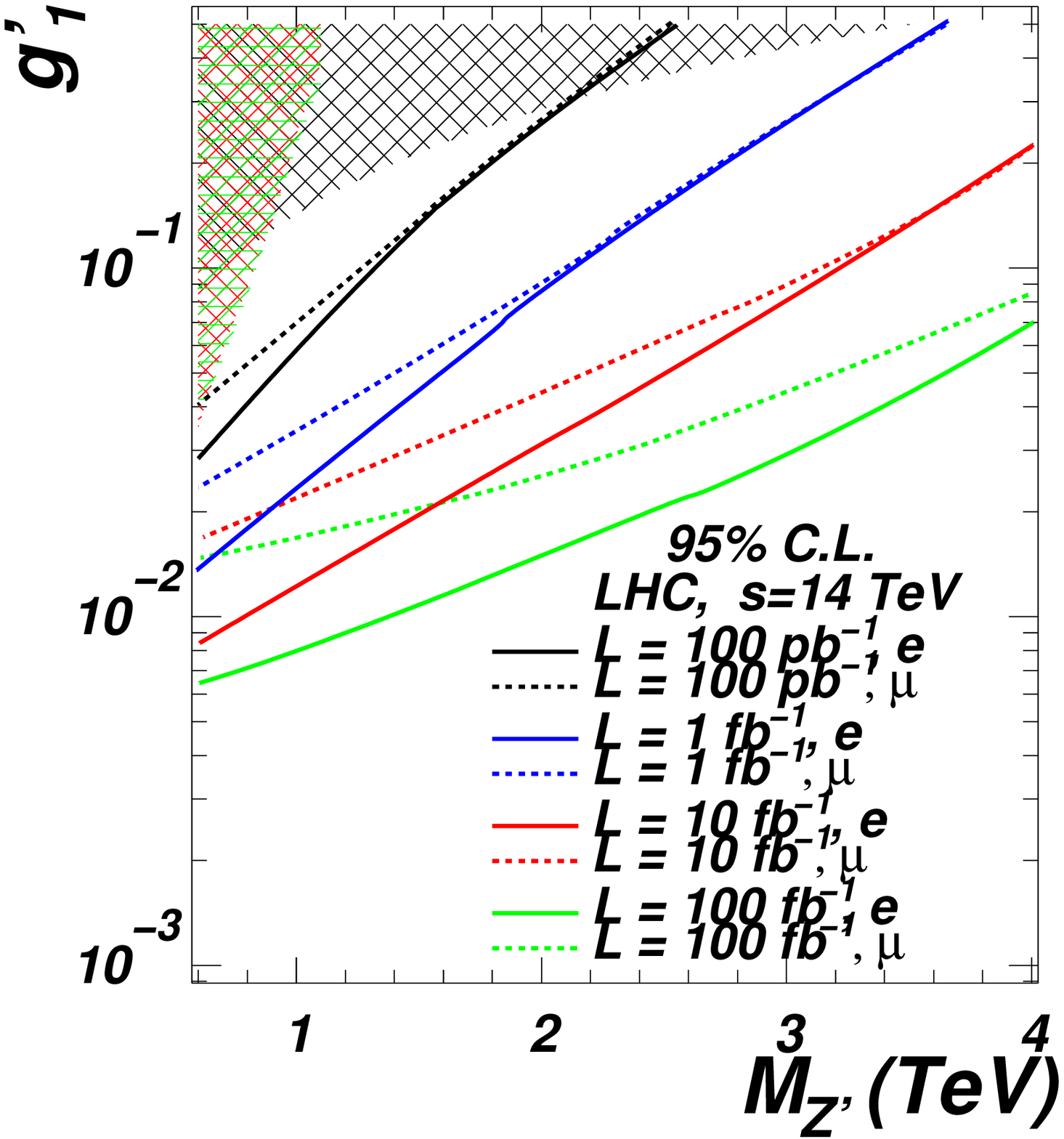}}
  \subfloat[]{
  \label{lumi14_excl}
  \includegraphics[angle=0,width=0.48\textwidth ]{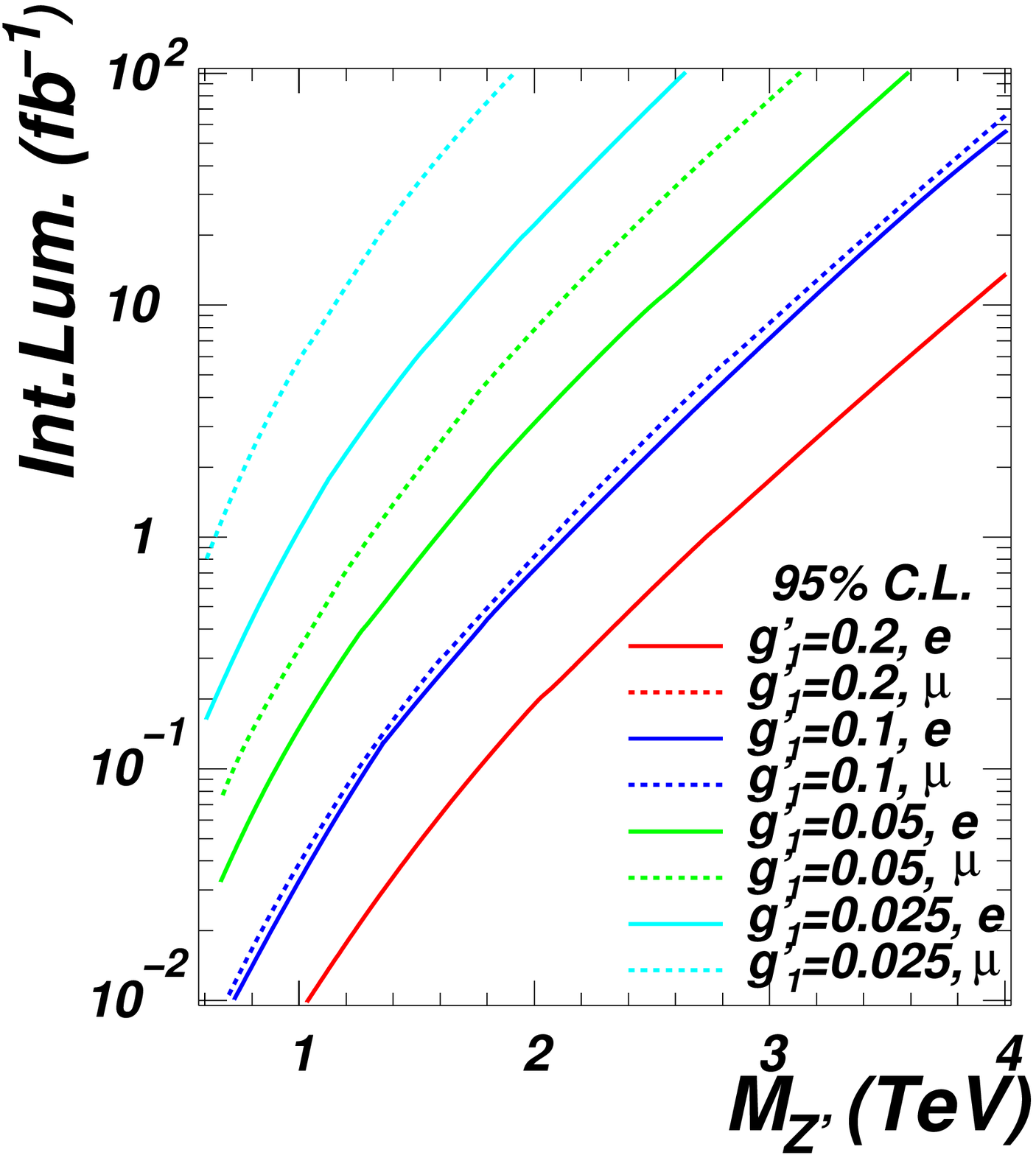}}
  \vspace*{-0.5cm}
  \caption{\it (\ref{contour14_excl}) Contour levels for $95\%$ C.L. plotted against $g'_1$ and $M_{Z'}$ at the LHC for selected integrated luminosities and (\ref{lumi14_excl}) integrated luminosity required for observation at $3\sigma$ and $5\sigma$ vs. $M_{Z'}$ for selected values of $g_1'$ (in which only the allowed combination of masses and couplings are shown), for  $\sqrt{s}=14$ TeV, for both electrons
and muons.
The shaded areas and the allowed $(M_{Z'},g'_1)$ shown are in accordance with eq.~(\ref{LEP_bound}) (LEP bounds, in black) and table~\ref{mzp-low_bound} (Tevatron bounds, in red for electrons and in green for muons).}
  \label{excl_14}
\end{figure}

Finally, we consider the LHC at $\sqrt{s}=14$ TeV. Due to the improved resolutions for both electrons and muons,
they have very similar exclusion powers for couplings $g'_1\gtrsim 0.1$, therefore setting similar constraints. Depending on the amount of data that will be collected, several maximum bounds can be set (see figure~\ref{contour14_excl}): i.e., with $10$ fb$^{-1}$ of data, the LHC at $14$ TeV can exclude at $95\%$ C.L. up to a mass of roughly $5$ TeV for a value of the coupling\footnote{This is the biggest allowed value for the consistency of the model up to a scale $Q=10^{15}$ GeV, from a Renormalisation Group (RG) analysis of the gauge sector of the model \cite{B-L,Basso:2010jm}.}
 $g'_1=0.4$. For $100$ fb$^{-1}$ and for the same value of the coupling, the LHC can exclude at $95\%$ C.L. masses up to roughly $6$ 
TeV. For $10$ fb$^{-1}$ it will be possible to exclude a $Z'_{B-L}$ boson for $M_{Z'}=600$ GeV if the coupling is greater than $1.8\,\cdot 10^{-2}(9\,\cdot 10^{-3})$ for muons(electrons), and values of the coupling greater than $1.5\,\cdot 10^{-2}(6.5\,\cdot 10^{-3})$ for an integrated luminosity of $100$ fb$^{-1}$.
Figure \ref{lumi14_excl} shows the integrated luminosity that is required to excluded a certain $Z'_{B-L}$ boson mass for fixed values of the coupling. As previously noticed, electrons and muons set the same limits for $g'_1 \geq 0.1$, a smaller value than for $\sqrt{s}=10$ TeV due to the improved resolutions. An integrated luminosity of $10$ fb$^{-1}$ is required to exclude a $Z'_{B-L}$ mass up to $3.8$ TeV for $g'_1=0.2$, instead $40$ fb$^{-1}$ reduces this to
$g'_1=0.1$.
For an integrated luminosity of $10$ fb$^{-1}$ the LHC experiments will be able to exclude the $Z'_{B-L}$ boson for masses up to $3.1$ TeV for $g'_1=0.1$, $2.5(2.1)$ TeV for $g'_1=0.05$ and $1.7(1.1)$ TeV for $g'_1=0.025$, when considering decays into electrons(muons). With $100$ fb$^{-1}$ of data, more stringent bounds can be derived: for $g'_1=0.05(0.025)$ the $Z'_{B-L}$ boson can be excluded for masses up to $3.6(2.6)$ TeV in the electron channel, and up to $3.1(1.9)$ TeV in the muon channel. 

The $95\%$ C.L. exclusions for the LHC at $\sqrt{s}=14$ TeV are summarised in  table~\ref{2sigma_at_14TeV}, for selected values of $Z'$ masses and couplings.
\begin{table}[h]
\begin{center}
\begin{tabular}{|c||c|c|c||c|c|c|}
\hline
$\sqrt{s}=14$ TeV & \multicolumn{3}{|c||}{$pp\rightarrow e^+ e^-$} & \multicolumn{3}{|c|}{$pp\rightarrow \mu^+ \mu^-$} \\
\hline
$g'_1$ & $M_{Z'}=1$ TeV & $M_{Z'}=2$ TeV  & $M_{Z'}=3$ TeV & $M_{Z'}=1$ TeV & $M_{Z'}=2$ TeV  & $M_{Z'}=3$ TeV \\
\hline
0.025 & 1.0   & 20   & $>$100	& 7    & $>$100 & $>$300    \\  
0.05  & 0.15  & 3    & 30	& 0.40 &  8     &   80    \\ 
0.1   & 0.04  & 0.7  & 7 	& 0.04 & 0.8    &   9    \\ 
0.2   & $-$   & 0.2  & 2  	& $-$  & 0.2    &   2    \\ 
\hline
\end{tabular}
\end{center}
\vskip -0.5cm
\caption{\it Minimum integrated luminosities (in fb$^{-1}$) for a $95\%$ C.L. exclusion for selected $Z'_{B-L}$ boson masses and $g_1'$ couplings in the $B-L$ model. No numbers are quoted for already excluded configurations.}
\label{2sigma_at_14TeV}
\end{table}

\section{Conclusions}\label{sect:conc}

We have presented the discovery potential for the $Z'$ gauge boson of the $B-L$ minimal extension of the SM at the LHC for CM energies of $\sqrt{s}=7$, $10$ and $14$ TeV, using the integrated luminosity expected at each stage. This has been done for both the $Z'_{B-L}\rightarrow e^+e^-$ and $Z'_{B-L}\rightarrow \mu ^+\mu ^-$ decay modes, and includes the most up-to-date constraints coming from LEP and the Tevatron.
The comparison of the (irreducible) backgrounds with the expected backgrounds for the D$\O$ experiment at the Tevatron validated our simulation. We proposed an alternative analysis that has the potential to improve sensitivities.
 We also looked in detail at the different resolutions, showing that electrons and muons present very similar discovery power for values of the coupling bigger than roughly $0.2(0.1)$, for $\sqrt{s}=10(14)$ TeV.

We are overall confident that the inclusion of further background, as well as a realistic detector simulation, will not have a considerable impact on the results we presented. In fact, as noted in section~\ref{sect:comput}, all detector effects can be casted in the form of a {\it signal acceptance}, including also the effect of kinematic and angular acceptance cuts. By looking at Refs.~\cite{Abazov:2010ti,Aaltonen:2011gp,Khachatryan:2010fa}, we estimate an overall acceptance factor of $\sim 70\%$, which we found to be approximately constant over the mass regions considered, to be applied to our parton level results [once the cuts of eqs.~(\ref{LHC_cut}) and (\ref{Tev_cut}) are considered]. This acceptance is mainly related to the leptons identification, both at the Tevatron and at the LHC. On the significance, as in eq.~(\ref{signif}), the reduction is then of $\sim 84\%$.

A general feature is that greater sensitivity to the $Z'_{B-L}$ resonance is provided by the electron channel. At the LHC this has better energy resolution than the muon channel. A further consequence of the better resolution of electrons is that an estimate of the gauge boson width would eventually be possible for smaller values of the $Z'_{B-L}$ boson mass than in the muon channel.
Limits from existing data imply that the first couplings that will start to be probed at the LHC are those around $g'_1=0.1$. Increased luminosity will enable both larger and smaller couplings to be probed.

Our comparison showed that, for integrated luminosity of $10$ fb$^{-1}$, the Tevatron is still competitive with the LHC in the electron channel and in the small mass region, being able to probe the coupling at the level of $5\sigma$ down to a value of $4.2 \cdot 10^{-2}$. The LHC will start to be competitive in such a region only for integrated luminosities close to $1$ fb$^{-1}$ at $\sqrt{s}=7$ TeV, or equivalently $500$ pb$^{-1}$ at $\sqrt{s}=10$ TeV (comparing the luminosities required for a $5\sigma$ discovery of $Z'_{B-L}$ boson of $600$ GeV mass for $g'_1=0.05$). Also, at $\sqrt{s}=7$ TeV the mass reach will be extended from the Tevatron value of $M_{Z'}=850$ GeV, with electrons, up to $1.25(1.20)$ TeV for electrons(muons).
The muon channel at Tevatron needs more than $10$ fb$^{-1}$ to start probing the $Z'_{B-L}$ at $3\sigma$. Hence, it has not been studied.

The LHC at $\sqrt{s}=10$ TeV will be able to further extend the kinematic reach of the $Z'_{B-L}$ boson, being able to probe it for masses much bigger than those available at the Tevatron,
 up to $M_{Z'}=1.8(1.7)$ TeV with $\int \mathcal{L}=1$ fb$^{-1}$ of data, depending on whether one is looking at electrons(muons).

When the data from the high energy runs at the LHC become available, the discovery reach of $Z'_{B-L}$ boson will be extended towards very high masses and small couplings in regions of parameter space well beyond the reach of the Tevatron and comparable in scope with those accessible at a future LC \cite{bbmp}.

If no evidence is found at any energies, $95\%$ C.L. limits can be derived, and, given their better resolution, the bounds from electrons will be more stringent than those from muons, especially at smaller masses.

While this work was in progress, other papers dealing with the discovery power at the LHC for the $Z'_{B-L}$ boson appeared, for CM energies of $7$, $10$ \cite{Salvioni:2009mt} and $14$ TeV \cite{Mine}, as well as for other popular $Z'$ boson models. Our results broadly agree with those therein.

\section*{ACKNOWLEDGEMENTS} 
\input{acknowledgements.tex}

\input{bibl.tex}
\end{document}

%% file: acknowledgements.tex
LB thanks Muge Karagoz Unel and Ian Tomalin for useful discussions.
SM is financially supported in part by 
the scheme `Visiting Professor - Azione D - Atto Integrativo tra la 
Regione Piemonte e gli Atenei Piemontesi'.
We thank the NExT Institute for financial support.